\documentclass[manuscript]{aastex}
\usepackage{outlines}
\usepackage{enumitem}
\usepackage{float}
\usepackage{amsmath}
\usepackage{siunitx}
\usepackage{mathtools}
\usepackage{graphicx}
\usepackage{color}
\begin{document}

\newcommand{\vdag}{(v)^\dagger}
\newcommand{\myemail}{gbbasri@berkeley.edu}

%% You can insert a short comment on the title page using the command below.

%\slugcomment{to appear in The Astrophysical Journal}

%% If you wish, you may supply running head information, although
%% this information may be modified by the editorial offices.
%% The left head contains a list of authors,
%% usually a maximum of three (otherwise use et al.).  The right
%% head is a modified title of up to roughly 44 characters.
%% Running heads will not print in the manuscript style.

\shorttitle{Starspot Lifetimes}
\shortauthors{Basri et al.}

%% This is the end of the preamble.  Indicate the beginning of the
%% paper itself with \begin{document}.

%\begin{document}


\title{A New Method for Estimating Starspot Lifetimes Based on Autocorrelation Functions}


\author{Gibor Basri, Tristan Streichenberger, Connor McWard, Lawrence Edmond IV, Joanne Tan, Minjoo Lee, Trey Melton}

\affil{Department of Astronomy, University of California,
    Berkeley, CA 94720}

\email{gbbasri@berkeley.edu} 

%%\and


%\altaffiltext{1}{University of California Berkeley.}

\begin{abstract}

We present a method that utilizes autocorrelation functions from long-term precision broadband differential light curves to estimate the average lifetimes of starspot groups for two large sample of Kepler stars: stars with and without previously known rotation periods. Our method is calibrated by comparing the strengths of the first few normalized autocorrelation peaks using ensembles of models that have various starspot lifetimes. We find that we must mix models of short and long lifetimes together (in heuristically determined ratios) to align the models with the Kepler data. Our fundamental result is that short starspot group lifetimes (1-4 rotations) are implied when the first normalized peak is weaker than about 0.4, long lifetimes (15 or greater) are implied when it is greater than about 0.7, and in between are the intermediate cases. Rotational lifetimes can be converted to physical lifetimes if the rotation period is known. Stars with shorter rotation periods tend to have longer rotational (but not physical) spot lifetimes, and cooler stars tend to have longer physical spot lifetimes than warmer stars with the same rotation period. The distributions of the physical lifetimes are lognormal for both samples and generally longer in the first sample. The shorter lifetimes in the stars without known periods probably explain why their periods are difficult to measure. Some stars exhibit longer than average physical starspot lifetimes; their percentage drops with increasing temperature from nearly half at 3000K to nearly zero for hotter than 6000K. 

\end{abstract}

%\begin{outline}[enumerate]

%\end{outline}


\keywords{starspots --- stars: magnetic field --- stars: activity --- stars: late-type}

\section{Introduction\label{sec:Intro}}

The Kepler mission \citep{Bor10} is unique in having observed nearly 200,000 stars over four years, sometimes almost continuously, with a half-hour cadence and a photometric precision measured in parts per million. Its primary purpose was to use transits to discover the statistics of exoplanets in inner planetary systems down to Earth-sized planets, and it was a stunning success on that mission. In service of this it collected the brightness variations of all the stars it was observing and this constitutes a dataset for a large number of stars with qualities that we previously only had for the Sun. Kepler inaugurated the field of statistical asteroseismology and also provided a large dataset on starspots. 

One of the easiest parameters that can be found by studying starspot light curves is stellar rotation periods, although it is not quite as easy for a lot of stars as one might think. Various techniques have been tried with similar levels of success \citep{Aig15}. In particular autocorrelation functions (ACFs) were employed by \citet{MMA14} (hereafter MMA14) to infer stellar rotation periods from Kepler light curves in what has become the standard reference for Kepler rotation periods. It should be noted that they were only able to determine confident periods for less than half the stars they tried. Other authors have tried to infer other stellar and starspot properties from the same light curves. Differential rotation measurements have been claimed by, among others, \citet{RRB13}, \citet{Rein15}, \citet{Das16}, and \citet{Sant17} (the latter paper also mentions several others). Activity cycles have been claimed by authors including \citet{Vida14}, \citet{Ark15}, \citet{Leht16}, \citet{Rein17}, \citet{Mont17}, and \citet{Niel18}; some of these papers also claim to see differential rotation. 

\cite{Basri20} (hereafter BS20) cast some doubt on the extent to which the light curves are revealing the sought-after effects, or are simply exhibiting similar behavior due to the random appearance and disappearance of spots in random positions. They did, however, find a more systematic behavior of light curves with starspot lifetime that showed real potential to provide more definitive information. Light curves generated with longer lifetimes show more periodic behavior and more systematic amplitude changes over time. Their models tested different numbers of spots on a star but kept the maximum size of individual spots fixed. \cite{Isik20} utilize a more physical representation of spots based on the solar model, but found that they have to artificially increase the frequency at which spots recurrently appear near their previous location in order to generate the sort of larger-amplitude more periodic light curves that many of the Kepler stars exhibit. They call this property ``nesting" (based on older solar terminology); it is similar in effect to the lifetime that is meant in stellar work and that BS20 implemented. More nesting is equivalent to longer lifetimes in a general sense (but not in detail).

Sunspots themselves exhibit lifetimes that depend on their size; this has been codified in the ``Gnevyshev–Waldmeier rule" that is a simple linear relation between maximum spot size and lifetime. \citet{Brad14} discuss the possible physics behind this relation, which has sometimes also been suggested to hold for certain stars. The mechanism of spot decay is thought to be dissolution of the concentrated collection of magnetic field due to dissipation, decay, and cancellation of the magnetic field caused by turbulent (often convective) motions of the surface plasma. They find that the size of supergranulation plays a strong role. Differential rotation could certainly also play a role if the shear is strong enough. The problem is that we don't understand stellar differential rotation very well, and it is unclear how the atmospheric turbulence or supergranulation responds to spotted regions that are much larger than ever seen on the Sun.

The determination of starspot sizes from light curves is also subject to some of the same criticisms BS20 leveled against the other attempted measurements. Differential light curve variations only reflect changes in the spot distribution asymmetries, not necessarily true spot coverage changes. Measurements made with Doppler imaging are less subject to that issue, but these are often of stars with large and nearly permanent polar spots. It is certainly the case that young active stars hold their spot patterns (light curve shapes) for many rotations, so it is likely that they indeed have longer spot lifetimes if we take that to mean the permanence of very large spotted regions. 

\citet{Gil17} (hereafter GCH17) proposed a method of using ACF peak heights to infer starspot lifetimes, since if a spot (group) lasts for several rotations that will strengthen the autocorrelation signal. They used an MCMC fit to the ACF to find a decay timescale from the decline in the heights of the first several ACF peaks. This was then related to the stellar rotation period, which they recomputed using a similar method to MMA14 (with some disagreements in the values found). They also examined the relation between the decay timescale and the photometric variability, finding that stars with larger variability (which they interpreted as due to larger spots) have longer lifetimes. GCH17 considered limited cases of particular small segments of rotation periods. They reached two basic conclusions, 1) larger starspots live longer on stars as well as the Sun, and 2) cooler stars tend to have longer-lived spots. This paper starts from a related conceptual approach and applies it to a much larger and more diverse sample of stars.

\section{Analysis \label{sec:analysis}}

\subsection{The Stellar Sample \label{sec:sample}}

The first sample of stars we consider is nearly the same as that published by MMA14. That group was selected by them from a larger sample of Kepler light curves as the ones for which they felt reasonably confident they had detected a rotation period. We have restricted it slightly to stars whose stellar parameters have been updated by Gaia, since the Kepler Input Catalog temperatures and stellar radii used by MMA14 are known to be somewhat inaccurate. To more cleanly separate main sequence (MS) stars from subgiants (SG) in this sample, we utilize the Gaia DR2 temperature and radius data for the $177,911$ Kepler stars presented by \citet{Berg18}. We use two stellar isochrone tables \citep{Spada17} to set a specific (if somewhat arbitrary) boundary between stars that are on the main sequence and stars have begun to sufficiently evolve off the main sequence on their way to becoming subgiants. 

\begin{figure}
\begin{center}
\includegraphics[width=\linewidth]{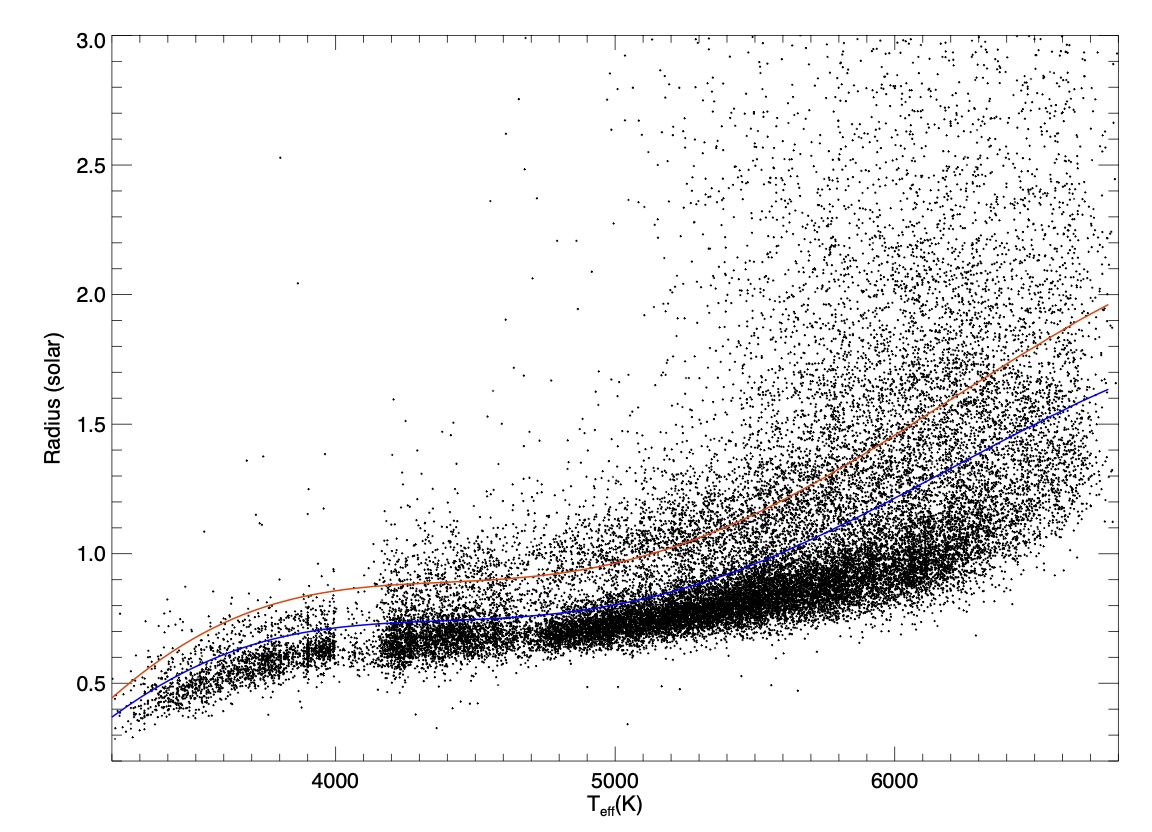}
\end{center}
\caption{The temperature-radius diagram for the MMA14 stars we analyze. The blue line is the top of our conservative main sequence and above the red line are the sub-giants. The NP21 sample is not shown but they all lie below the blue line. }
\label{fig:HRD}
\end{figure}

The two isochrone tables include temperature, radii, and age data for stars with masses between $0.3M_{\odot} -  1.0M_{\odot}$ and $0.6M_{\odot} - 3.0M_{\odot}$. We first determined the median age for stars at each mass in the tables. There were slight differences in the overlap region between the two isochrone tables; we averaged those two values to end up with a set of median model ages of stars with masses between $0.3M_{\odot} - 3.0M_{\odot}$. We then set a rather conservative upper boundary condition below which stars are confidently on the main sequence as the radii of stars with ages $1.2$ times greater than the median model ages for their mass. This translates to the Sun having expanded by 10\% from its current size. The SG sample extends up to the Gaia radii of the largest stars in the MMA14 sample. Finally, we used the Gaia data along with the list of known Kepler binaries \citep{Kirk16} to remove binaries from our sample. About 4000 stars were in the MMA14 sample but not listed in by \citet{Berg18} preventing classification of their MS status by our method. This resulted in our final MMA14 comparison sample of $30,050$ stars shown in Figure \ref{fig:HRD} with temperatures between 3200K and 6800K, $23,750$ of which are MS and $6300$ are SG. The red line marks the top of our nominal main sequence at 1.2 times the radii of the conservative main sequence. It turned out that our results are not very different for the SG compared with the MS so we usually just refer to the whole sample. 

We also studied another large sample of stars, namely the Kepler stars with Gaia parameters in \citet{Berg18} that place them on our conservative definition of the main sequence that were not assigned a rotation period by MMA14 (but are in the longer list of stars they attempted). This set contains $32,489$ MS stars with similar temperatures to the MMA14 sample (up to 6500K instead of 6800K) that we dub the NP21 sample (for ``non-periodic" or ``not published"). Because of the biases involved in finding rotation periods from the Kepler light curves, it is likely that the NP21 stars are the ``missing" older solar-type and cooler stars that are older than the Sun and not active enough to readily reveal their rotation periods via starspot modulation. Our conservative cut in radius ensures this sample really only contains main sequence stars.

\subsection{Conditioning the Kepler Data \label{sec:condata}}

The Kepler reduction pipeline underwent continual improvement from launch until the final product in 2016 (called DR-25). Its purpose is to convert the raw pixel intensities for each target into a calibrated intensity that could be added to a light curve. Analyses done before the last data release used whichever version was last available, so the light curves from DR-25 are not necessarily identical to those used in earlier papers (including MMA14). In addition we have re-conditioned the light curves in a number of ways for this paper that we hope helps with the final study of autocorrelation functions. We now describe those steps.

After obtaining the quarterly light curves from the MAST we first applied a median filter on a few hour timescale with the intent of removing planetary transits, flares, cosmic rays, and other sharp glitches. Obviously this is counter to the main purpose of the Kepler mission but it helps with the starspot analysis. We next replaced all points on the intensity vector that don't have valid data with zeroes so that they are not counted during the calculation of the autocorrelation function. This includes quarters with no data at all (which we kept a catalog of) or those with fewer than 100 cadences, and times when the spacecraft was not on target due to data downloads or anomalies of various types. We also decided not to include the short quarters 1 and 17. We then binned the light curves by a factor of ten in time and interpolated them onto a single continuous time vector; each time point is about 0.2 days long. Starspots don't influence the light curve on timescales shorter than that and the noise is substantially reduced along with data storage and processing requirements. Each quarter is normalized so its median is zero and the units of differential intensity are parts per thousand (ppt). We do not de-trend the quarters with a low-order polynomial for this project. 

The next filter is for quarters that have anomalously large ranges or few zero-crossings. The range and zero-crossing metrics were introduced by \cite{Basri11} and have been utilized by a number of authors. Range is simply a measure of the amplitude of variability of the normalized differential light curve over a quarter; it is the difference between the 95th percentile and the 5th percentile brightest intensities. The choice to exclude the ten percent most extreme points is somewhat arbitrary but driven by experience on the Kepler dataset. The zero-crossing metric is primarily a way of distinguishing light curves that have large slow excursions around the median (few zero crossings) from those with relatively faster excursions well above the noise (moderately many zero crossings) and those that are just noisy (many zero crossings). After computing these metrics for all non-zero quarters in a light curve, quarters are marked as ``reject" if their range is greater than 5 times the median range or the number of zero-crossings is less than a third of the median of those. This procedure removes what turns out to be a significant number of quarters that have some sort of data reduction problem. These problems often manifested as a few large excursions in the quarter while the rest of the quarters look qualitatively different. Leaving such anomalies in the light curve can significantly influence the autocorrelation function.

\subsection{Autocorrelation Function Diagnostics \label{sec:ACFdiag}}

Because MMA14 utilized an autocorrelation technique to infer the periods, their sample naturally contains ACFs that have a series of peaks with a reasonably regular spacing. That does not mean that all the ACFs are fully regular either in peak spacing or peak strengths -- many are not. They may contain peaks that don't fit onto the harmonic sequence in time or be missing such peaks. The general expectation that the first harmonic peak will be the highest (meaning that the correlation is greatest after one rotation) can also be violated. As the regularity decreases the confidence in the inferred rotation period decreases. MMA14 had criteria for deciding when their confidence fell too low; these criteria sometimes involved visual inspection and some subjectivity. We will re-examine whether the ACF method for finding rotation periods can be improved in a future paper but for now we accept the MMA14 rotation periods as is.

We now describe how we arrive at our final ACF and its diagnostic information. We first clip the light curve so that all the points lie within its range as calculated above. This means that the 10\% most extreme positive and negative points are flattened to the high or low values of the differential intensity that set the range. This is to prevent really extreme points from influencing the ACF. We next check to see if the range is so small that smoothing the light curve is desirable. This is because we found that light curves that are mostly noise can generate spurious ACF peaks that are suppressed with some pre-smoothing. The act of performing the autocorrelation is of course itself a very significant form of smoothing, so this check is only applied for ranges that are less than 10 ppt. When desired we apply a boxcar smoothing with a width of the reciprocal of the range times ten and not more than ten cadences wide.

%The autocorrelation itself is applied over some span of time (a certain number of time points in each direction). Because the light curve is being shifted over itself and one doesn't know what it looks like when shifted off either end, the shift cannot be too large a fraction of the total duration without generating results that might be different if the light curve had additional time coverage. Our procedure is confined to shifting of known points. We shift the inner two-thirds of the light curve one-third of the way in both directions out to the first and last times. Because the autocorrelation is the product of the light curve with its shifted self, any zeros in the data produce a zero product whether shifted or not. Note that our procedure is not exactly the same as what is done for a standard autocorrelation. In that case the whole light curve is shifted against itself, with zeroes being added on the end as the shifted curve extends beyond valid data. This is done in both directions, producing a symmetric ACF. In the procedure used here the ACF is not symmetric because only valid points are allowed in both directions, but the portion of the curve near the beginning is not the same as the portion of the curve near the end. After computing the initial ACF, we fold it around zero shift to obtain an average ACF. 

A standard autocorrelation function is generated by shifting the whole curve over itself in both directions. Because this produces an ACF that is symmetric about zero shift, we only use the positive shift side. This is then normalized by subtracting its minimum then dividing by the resulting maximum, so the ACF is unity at zero shift and zero at some larger shift (often but not necessarily the first ACF minimum). Next we locate all the minima (dips) and maxima (peaks) in this normalized ACF. Their locations are then converted from vector index to time (in days) and the normalized heights and time locations of all the peaks and dips stored for analysis. 

The number of peaks found depends on the rotation period and the extent to which the ACF is regular. Some ACFs also contain peaks at fractional harmonics of the period (especially half-periods).  We kept at most the first 20 peaks (and one extra dip so the last peak was well-defined). Because our methodology repeatedly refers to peaks by number (first peak, second peak, etc.) we next took some care to remove the peaks that are not indicative of rotational modulation. Some are unrealistically near each other or have amplitudes that are too small. Such peaks occur when the light curve is too noisy, the starspot signal is too aperiodic, or the missing or anomalous quarters introduce them. 

\begin{figure}
\begin{center}
\includegraphics[width=\linewidth]{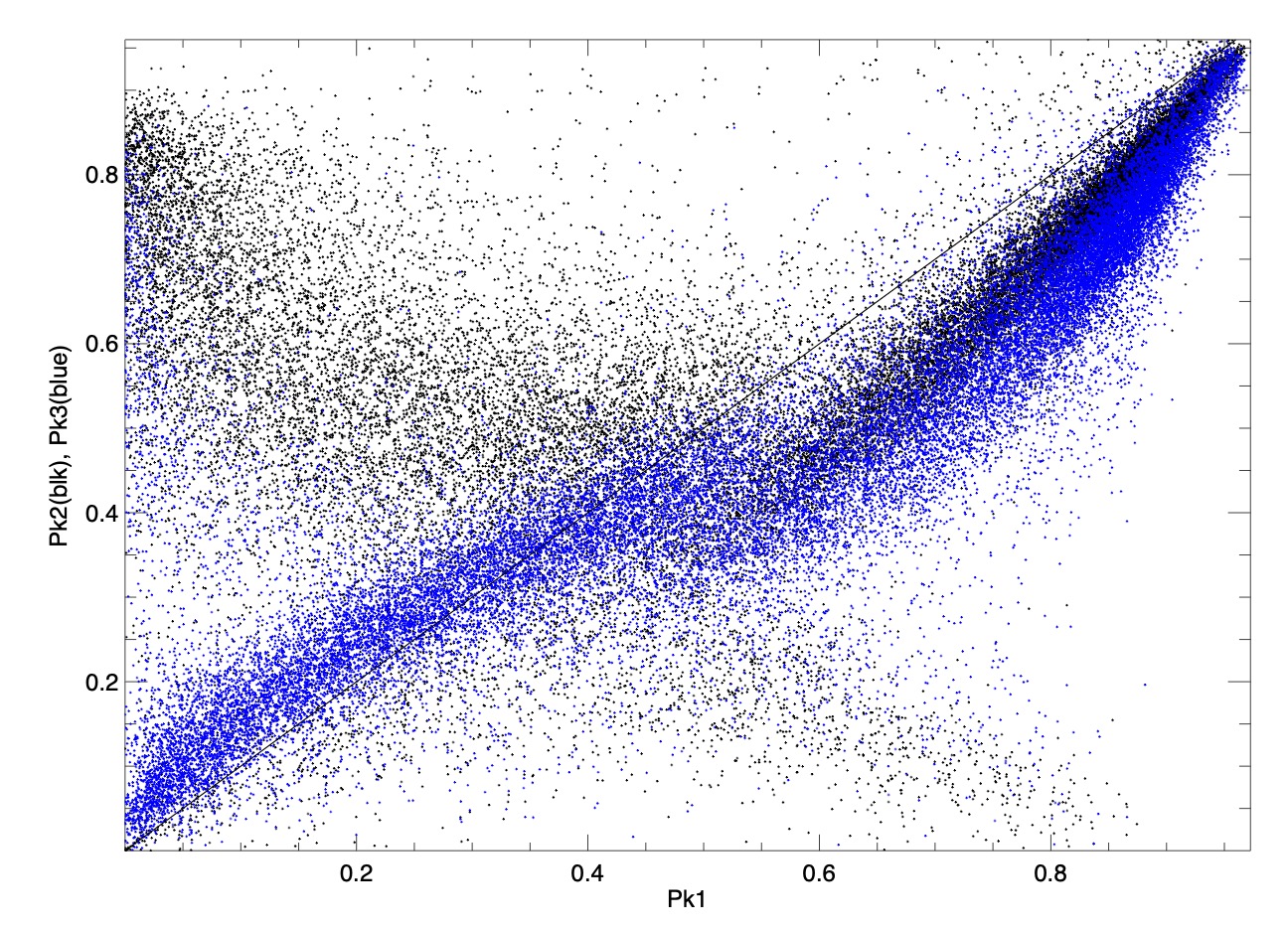}
\end{center}
\caption{The relation between ACF peak heights for the MMA14 sample of Kepler light curves. The abscissa is the height of the first peak, while on the ordinate black points give the height of the second peak and blue points give the height of the third peak. The diagonal black line indicates equality between the peaks. }
\label{fig:McQACF}
\end{figure}

We initially characterized peaks with three possible strengths: the height of the normalized ACF maximum there, the difference between that and the average of the adjacent minima (which we call depth), or the integrated area above the adjacent minima (which we call area). Our first filter removed peaks whose depth was less than 0.03 in our normalized units. Occasionally one also finds two peaks very close to each other (essentially, a tiny spurious dip within a peak) so we removed the second of such pairs when they were closer than an eighth of the mean peak spacing. After this cleaning we kept only the first 5 of the peaks (not counting the central peak) and first 6 of the dips for further analysis. The peak counts refer to these remaining peaks, and we utilize just the absolute height of the normalized peaks as our diagnostic from here on. Our method differs from the method of GCH17 in that we do not try to fit the decay of the peak heights but only consider the relative heights of pairs of them.

Figure \ref{fig:McQACF} shows the distribution of the heights of the second and third peaks relative to the first peak. As expected the height of the second peak (black) is often closer to but smaller than the height of the first peak, and the height of the third peak is smaller relative to that. However this is only generally the case when the height of the first peak is about 0.5 or greater. Below that level, the second peaks are often found above the first peak, and this becomes more pronounced as the first peak height gets smaller. One reason for this is that as MMA14 realized and \cite{Basri18} elaborated, sometimes the half period generates an initial ACF peak but the rotation period is represented by the second peak with a greater height. The black points on the left half of the plot that lie well above the line are such cases. In those cases the third period will tend to resemble the first period as can be seen in Fig.\ref{fig:McQACF} for the blue dots. Somewhat unexpectedly the blue dots lie a bit above the equality line for lower values of the first peak, meaning that the third peaks are typically a little stronger than the first when the first peaks are weak enough. Of course these ACFs are not very strong in normalized terms; these are the more aperiodic cases.

\begin{figure}
\begin{center}
\includegraphics[width=\linewidth]{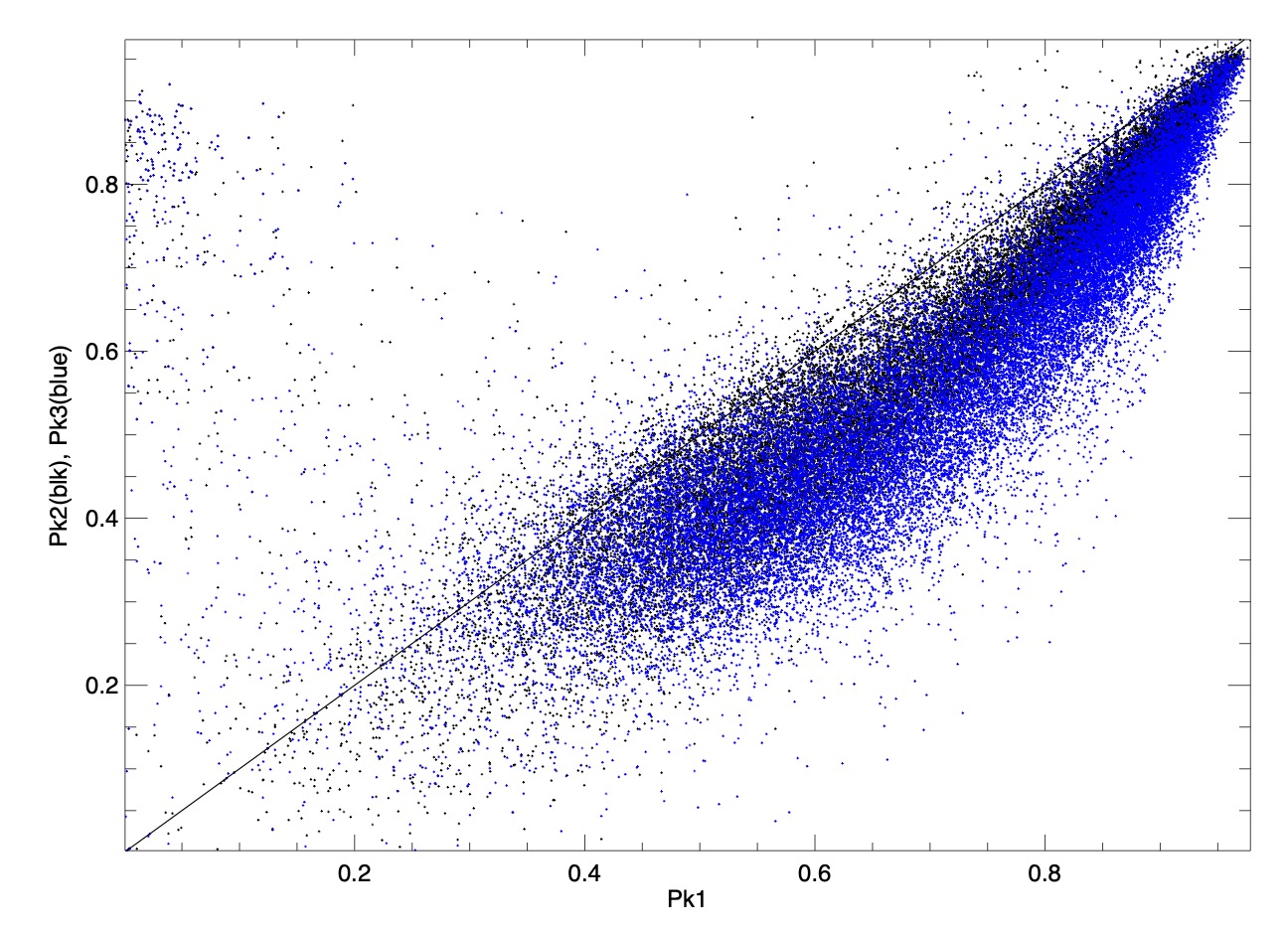}
\end{center}
\caption{The relation between ACH peak heights for the MMA14 sample of Kepler light curves. This is the same as the previous figure but uses the harmonic peak heights. Some black points are hidden beneath blue points. }
\label{fig:McQACH}
\end{figure}

To use the ACF to learn about starspot lifetimes one is primarily interested in the peaks that are diagnostic of the rotational longevity of spot distribution features. We found that for that purpose it makes better sense to modify the peak count so that it tries to count only peaks near the integral harmonics of the rotation period. This is an easy task in the case of models (where the rotation period is known) and for Kepler light curves where there is little question about the rotation period; one can ignore the half-period peaks when present in those cases. Unfortunately there are also many light curves for which the rotation period is less clear or ambiguous. We defer the problem of how this might affect results by adopting the MMA14 rotation periods as known, but return to it at the beginning of \S \ref{sec:LifeResults}. For now we implement a modified version of the ACF filtering that concentrates on integral harmonics of the specified period. We will designate these as ACH peaks. Some ACFs don't show a peak near where a harmonic is expected; we set such ACH peak heights to zero. This is similar to what GCH17 did when they ignored the ``interpulse" peaks.

Figure \ref{fig:McQACH} shows that this procedure does a lot to clean up the peak-peak relations. Not many of the rotational first peaks have heights less than about 0.4, and the second and third peaks are almost all smaller and often in descending order. The small set of points near the equality line represent cases where the procedures of MMA14 have the most difficulty distinguishing between the half and full rotation period because the ACFs have nearly the same strength in both cases. Examining the behavior of stars or models in a diagram like this is the basis of the rest of this paper so we will refer to the diagnostic represented by Fig.\ref{fig:McQACH} as a peak-peak plot (PPP) from now on. We will call the peak associated with the rotation period Pk1, and the next two harmonics Pk2 and Pk3.

\subsection{Peak Height Relations in Models \label{sec:Samples}}

In order to measure spot lifetimes from PPPs we have to understand how spot lifetime influences the locus of each instance in that space. The very concept of spot lifetime is not sharply defined; on the Sun spots last for different lengths of time depending on their size and their context within a spot group. BS20 conducted an extensive analysis of the formation of light curves in a multi-dimensional parameter space; refer to that paper for details of the modeling procedures and results. Among the primary parameters of their models are spot size, number, and lifetime, and stellar inclination. A given run has starspots with a fixed maximum radius in degrees and fixed lifetime measured in rotation periods ($L_{rot}$). Each ``spot" (which more appropriately represents a spot group) starts at a specified time and grows linearly to its maximum size over half its lifetime, then decays linearly. For each parameter set hundreds or thousands of runs were computed with most parameters fixed but spot locations and appearance times randomized. All the runs contained 50 rotation periods with 30 points in the light curve for each period.

BS20 found that one of the most easily discerned qualitative differences between the light curves from various model parameters is the spot lifetime. Simply by looking at the light curve one can guess whether it is on the longer ($L_{rot}>10$) or shorter ($L_{rot}<3$) end of the range of lifetimes. The longer the lifetime the more periodic the light curve looks, which suggests that the ACF could be a useful tool for extracting more precise information about the spot lifetime. Of course, stars do not behave in quite so convenient a manner (fixed lifetimes or spot sizes, for example) so one might expect some sophistication could certainly be added and some uncertainty is unavoidable. The models implement lifetimes through a linear growth and decay scheme, but we found that light curve properties were not very sensitive to exactly what scheme was employed. It will be interesting to see if more sophisticated modeling changes the relation between lifetime and our diagnostics derived below.

\begin{figure}[H]
\begin{center}
\includegraphics[width=\linewidth]{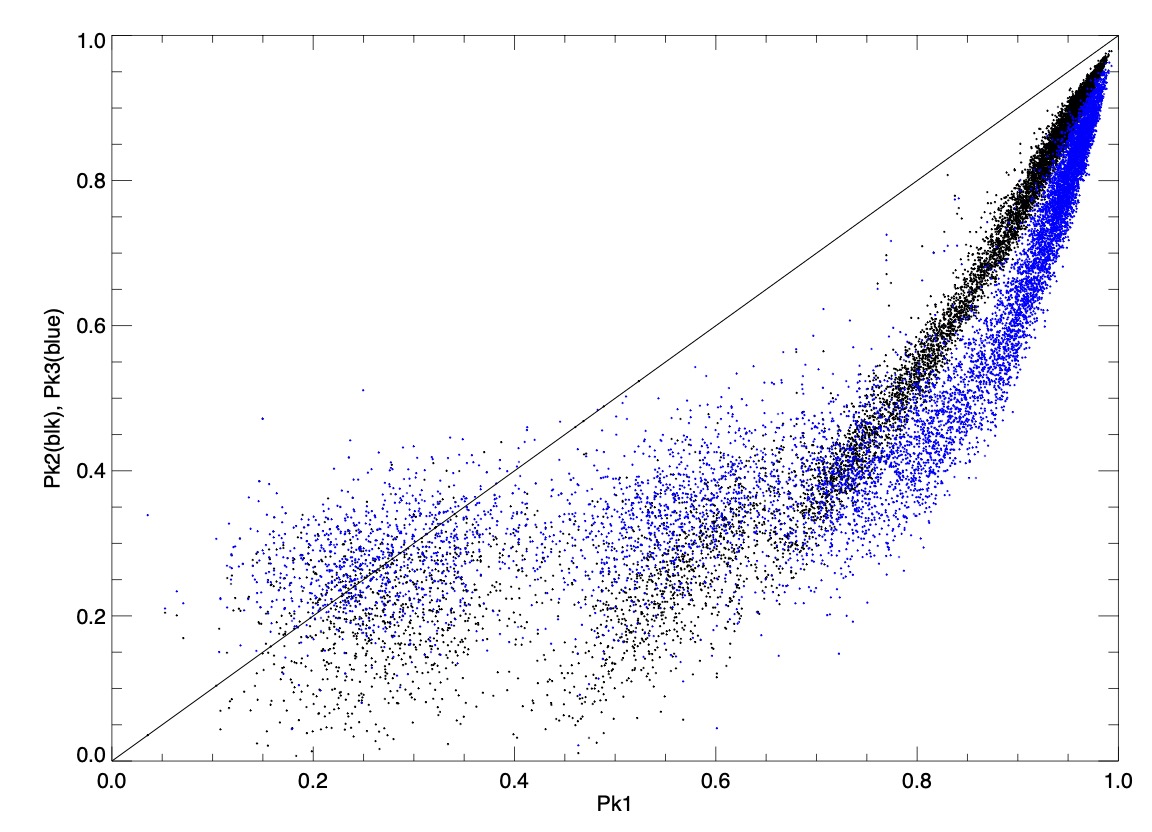}
\end{center}
\caption{The relation between ACH peak heights for the model light curves. This is the same as Fig.\ref{fig:McQACH} except the points come from a set of models with spot lifetimes ranging from 1 to 20 rotations rather than from real data. Blue points are plotted on top of black points and hide some of them. }
\label{fig:mod0pkh}
\end{figure}

A goal of this paper is to make a more comprehensive attempt to understand how light curves with a variety of lifetimes translate into locations in a PPP. To that end, we took a number of the models from BS20 and augmented them with other $L_{rot}$ as needed. For parsimony, employing the understanding gained about how the parameters affect light curves, we used model runs of 1000 trials each with the stellar inclination fixed at 60 degrees with fixed spot size and the spot number fixed at 6. Spots were given full latitude (and longitude) about where to appear. The only parameter explored in detail was $L_{rot}$, for which we computed a denser grid ranging from 0.5 to 50 rotations. Before using the model light curves to generate an autocorrelation function, we ``keplerized" them. This means that they were converted from absolute intensity light curves to differential intensity light curves (as Kepler gathers). The length of a ``quarter" for this process is 10 rotations; the light curve is flattened by a low-order polynomial within the quarter then median subtracted and put in units of parts per thousand so that it is similar to the Kepler data. This removes slow features that are due to large changes in total spot coverage over long time spans that Kepler would not be able to detect. 

Figure \ref{fig:mod0pkh} shows model results comparable to those for the Kepler ACHs in Fig.\ref{fig:McQACH}. These results are shown for models that are noiseless with a set of values for $L_{rot}$ from 2 to 50. It is apparent that although the model PPP looks somewhat similar to the data there are clear differences. The separation between the Pk1-Pk2 points (black) and Pk1-Pk3 points (blue) is cleaner and the relations are tighter. Pk3 is almost always smaller than Pk2 for high values of Pk1. This is what GCH17 had in mind in their analysis when they made an exploratory attempt to use the ACF to extract starspot lifetimes. In that study they restricted themselves to small controlled subsets of the MMA14 sample. In the upper right part of this PPP it is also clear that Pk2 and Pk3 decrease more rapidly compared with Pk1 for the models than for the observations. The model points become mixed together and spread out rapidly as one moves to values of Pk1 weaker than about 0.7, however.

\begin{figure}[H]
\begin{center}
\includegraphics[width=\linewidth]{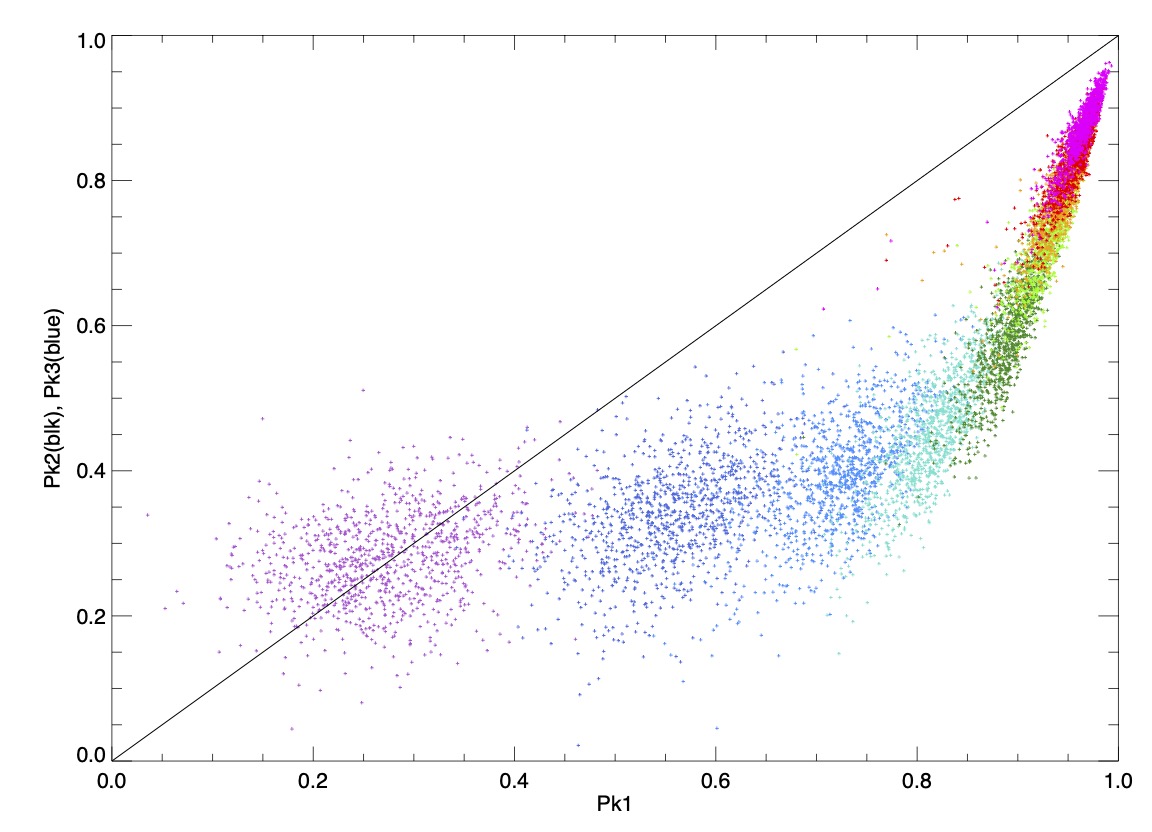}
\end{center}
\caption{This is the same as Fig.\ref{fig:mod0pkh} except the points are just the blue ones in that figure, but now colored by the lifetime each model used to generate them. The color code for each lifetime is: 2-purple, 3-dark blue, 4-light blue, 5-turquoise, 7.5-dark green, 10-light green, 12.5-orange, 15-red, and 20-magenta. }
\label{fig:lifeach}
\end{figure}

Figure \ref{fig:lifeach} shows how one might infer spot lifetimes in a PPP by separating out the points arising from models with different lifetimes by color. The grid of $L_{rot}$ shown is 2,3,4,5,7.5,10, 12.5, 15, and 20 rotations. It is clear that longer lifetimes are only found above Pk1 values of 0.7 and are squeezed into a narrow range in the PPP with Pk3. The shorter lifetimes occupy most of the PPP but there is still some separation between the shortest $L_{rot}$ (1-2 rotations) and $L_{rot}$ of 3-5. This holds the promise of at least being able to sort stars into short, medium and long $L_{rot}$. One can translate values of $L_{rot}$ into physical lifetime values $L_{day}$ for a given case if the rotation period is known in days, but that is not relevant for the models themselves. We will return to this later when interpreting the results for the real stars. 

The PPP in Fig.\ref{fig:lifeach} is not quite like that for the observed blue points in Fig.\ref{fig:McQACH}. The distribution for real stars stays closer to the equality line and is thicker in the upper right portion, and the models extend to lower values of Pk1. We therefore investigated whether the models could be modified in some way to better match the observations. If we can attain similar distributions in the PPP it becomes possible to assign values of $L_{rot}$ to real stars based on the model points, although it is clear they will come with some uncertainty. Our first experiment was to add noise to the models. This actually did work in the sense of making the model and data PPP look more similar, but the noise levels required are clearly significantly greater than are present in the data. 

Because the Sun has a mix of spot lifetimes we next tried adding a component of short lifetime ($L_{rot}=1$) light curves to each of the sets of light curves from the other values of $L_{rot}$. This was done by adding two random light curves together, one with each value of $L_{rot}$, in fractional amounts that we systematically varied looking for the optimal mix. Because each curve contains light deficits due to spots of a certain size, their amplitudes can be reduced by a constant factor to simulate the result that would accrue if the spots were simply smaller. Thus our procedure simulates cases where larger spots with longer lifetimes coexist with smaller shorter-lived spots. 

The best choice of fractional contributions in each case was decided by which combination produced points in the PPP that most closely matched the distribution in the Kepler PPP in the region where those points ended up. We tried adding the light curves for $L_{rot}=1$ in fractional amounts of 0.1,0.25,0.4,0.55,0.7, and 0.85 to each longer-lifetime light curve. After some experimentation we ended up with the following choices: for $L_{rot} < 5$ we used 0.85, for $L_{rot} = 5$ we used 0.7, for $L_{rot} = 7.5, 10$ we used 0.55, for $L_{rot} = 12.5$ we used 0.4, for $L_{rot} = 15, 20$ we used 0.25, and for $L_{rot} = 50$ we used 0.10. Although these are derived heuristically they have the property of possessing fractionally more short-lived spots in cases where the other spots are not too long-lived, and fractionally fewer when the other spots are more long-lived. We refer to the combined models by the longer of the values of $L_{rot}$ they contain. 

\begin{figure}[H]
\begin{center}
\includegraphics[width=\linewidth]{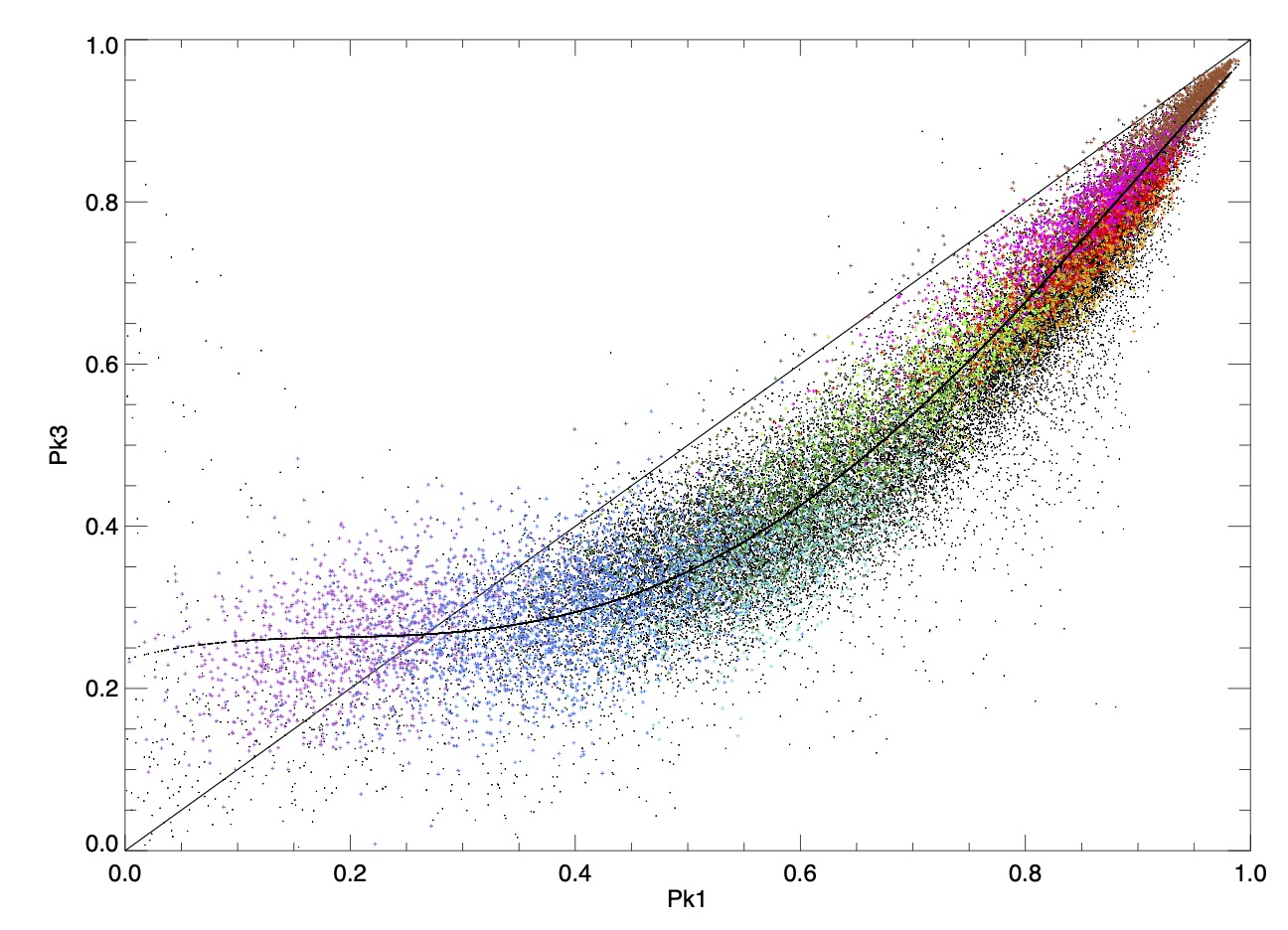}
\end{center}
\caption{The PPP for ACH peaks from combination models. The colors have the same meaning as in Fig.\ref{fig:lifeach} (with the addition of $L_{rot} = 50$ - brown), but now for models for which $L_{rot}$=1 and the other values of $L_{rot}$ have been combined in different fractional ratios (see text). The MMA14 points are shown with black dots; the model combinations were chosen to match them fairly well. The curved black line is a third-order polynomial fit through the MMA14 points. }
\label{fig:combppp}
\end{figure}

Fig.\ref{fig:combppp} shows that this procedure was fairly successful. The Kepler PPP is shown with black dots and the colored model points lie in the same region. The behavior of the two peak strengths thus now exhibits the same relations between the models as in the data and the range and dispersion of the points is also similar. We do not claim that this is the only way to make models fit this form of data, or that our models contain all the right physical features (certainly they do not since they use fixed inclinations and spot sizes). However this scheme reproduces the essence of the Kepler PPP by using models with mixes of different lifetimes. We also performed this analysis with Pk1/Pk2, but that did not provide much extra unique information. As seen in Fig.\ref{fig:McQACH} they are qualitatively similar but Pk2 is less different from Pk1 than Pk3 is.

\subsection{Methodology for Determining Rotational Spot Lifetimes  \label{sec:Method}}

Given the results of the last section, we turn to the question of how best to extract spot lifetime information from the distribution of model points compared with the distribution of observational points in a PPP. It is clear in Fig.\ref{fig:combppp} that there is not a one-to-one relation between the lifetime and the location in the PPP; there is a fair amount of mixing at a given value of Pk1. It is also clear, however, that lifetime increases in a general way as Pk1 increases and therefore that there is useful information in the PPP. We tried a few methods of extracting this information. In the end, we found three different methods as described below that agree fairly well with each other, and their combination produces the final set of lifetimes we ascribe to the Kepler stars.

\subsubsection{Peak Height Distributions \label{sec:PkDistr}}

\begin{figure}
\begin{center}
\includegraphics[width=\linewidth]{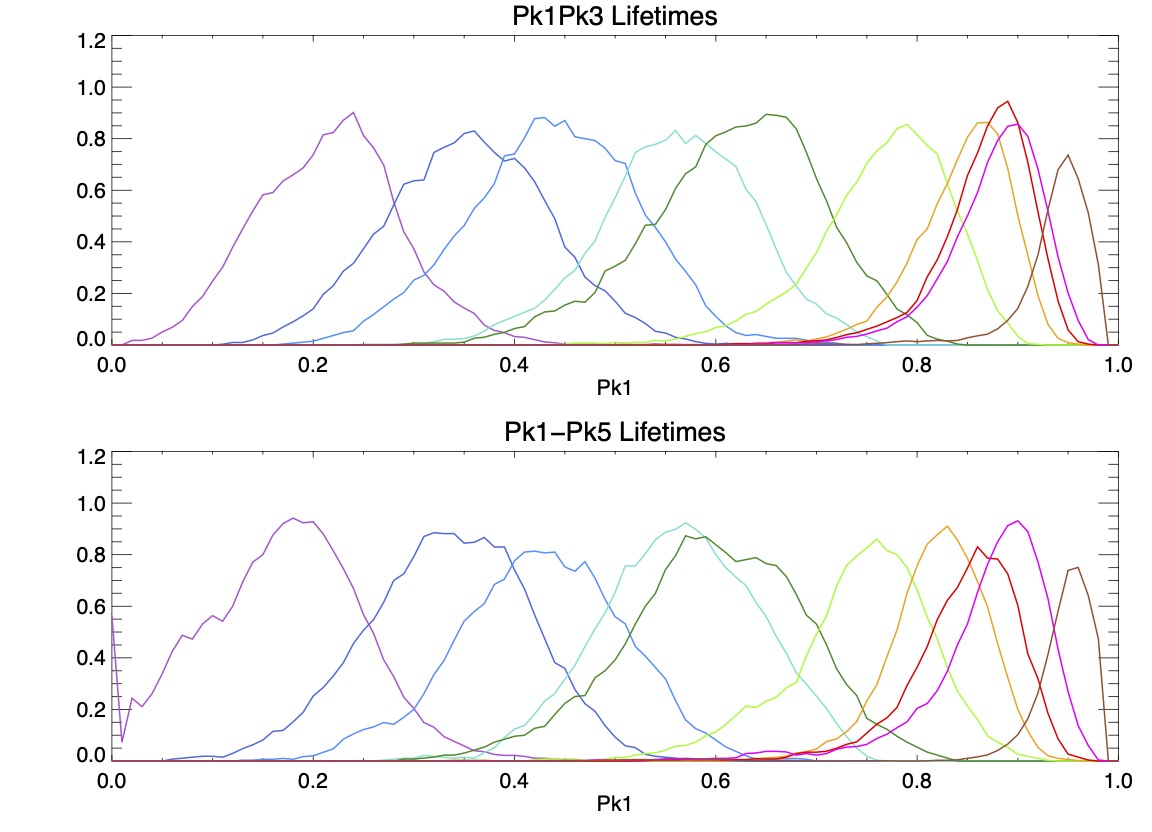}
\end{center}
\caption{The distributions of the models for each lifetime in the two cases considered. The abscissas are the Pk1 values closest to the observational fit curve. The upper panel is for the Pk1 vs Pk3 case and the lower panel is for the Pk1 vs Pk1-Pk5 case. The colors are for different lifetimes increasing from left to right with the same values as in Fig.\ref{fig:lifeach}. There are differences in the mean value of Pk1 implied for each lifetime between the two cases but they are not large. The lower panel separates the long lifetimes a little more effectively. }
\label{fig:lifeprob}
\end{figure}

For our first method, we began with the fit through the Kepler points (thick black line) in Fig.\ref{fig:combppp}, which comes from a third-order polynomial fit to Pk1 vs the observed points. We compute the shortest distance between this fit and each of the model points to assign a value Pk1' to each model point. For a given model point, Pk1' is the value on the abscissa of the perpendicular projection of the model's Pk1 onto the fit. Because the points are fairly close to the fit and the spread is smaller where the fit becomes more vertical, the values of Pk1' are fairly close to the values of Pk1. Each lifetime model produces a set of Pk1' values; their distributions are shown in the upper panel of Fig.\ref{fig:lifeprob}. Figure \ref{fig:lifeprob} makes several things clearer. When Pk1 is less than about 0.3, $L_{rot}$ is 2 or less. The region between 0.3 and 0.5 is occupied primarily by $L_{rot}$ between 3 and 4. Above about 0.8 there is a mix of lifetimes 10 and above. For Pk1 above 0.95 $L_{rot}$ is greater than 20 and it is difficult to know by how much. These results suggest a division of lifetime groups as follows: below Pk1 = 0.5 then $L_{rot}\leq 4$ (short), for Pk1 = 0.5-0.75 then $5\geq L_{rot}\leq 10$ (moderate), and for larger Pk1 values then $L_{rot}>10$ (long). We use the values of the individual peaks as the abscissa of a relation with lifetimes as the ordinate to find a polynomial relation between lifetime and Pk1'. Because the lifetimes extend from 2 to 50 with a small range of Pk1 above 10, we found it expedient to calculate the polynomial using the logarithm of the lifetime. 

This method produces a set of lifetimes (in units of rotation period) inferred from a set of values of Pk1' that can be applied to either model or observed values. When applying it we used the Pk1' values computed for the observed points. Since the actual Pk1 values from random instances of models for a given lifetime have a distribution, the lifetimes produced from this polynomial are only indicative within a range that itself depends on lifetime. They are mostly useful in a statistical sense, giving a rough idea of what $L_{rot}$ is likely to be near. While we could assign more specific probability or uncertainty distributions to these fitted lifetimes, the extra precision would be illusory since it is also not clear how well the models really reproduce the stellar realities except in a general way. That is, there must be systematic errors whose size is somewhat unknown but could easily be larger than the internal errors.

\subsubsection{Peak Difference Distributions \label{sec:PkDiff}}

\begin{figure}
\begin{center}
\includegraphics[width=\linewidth]{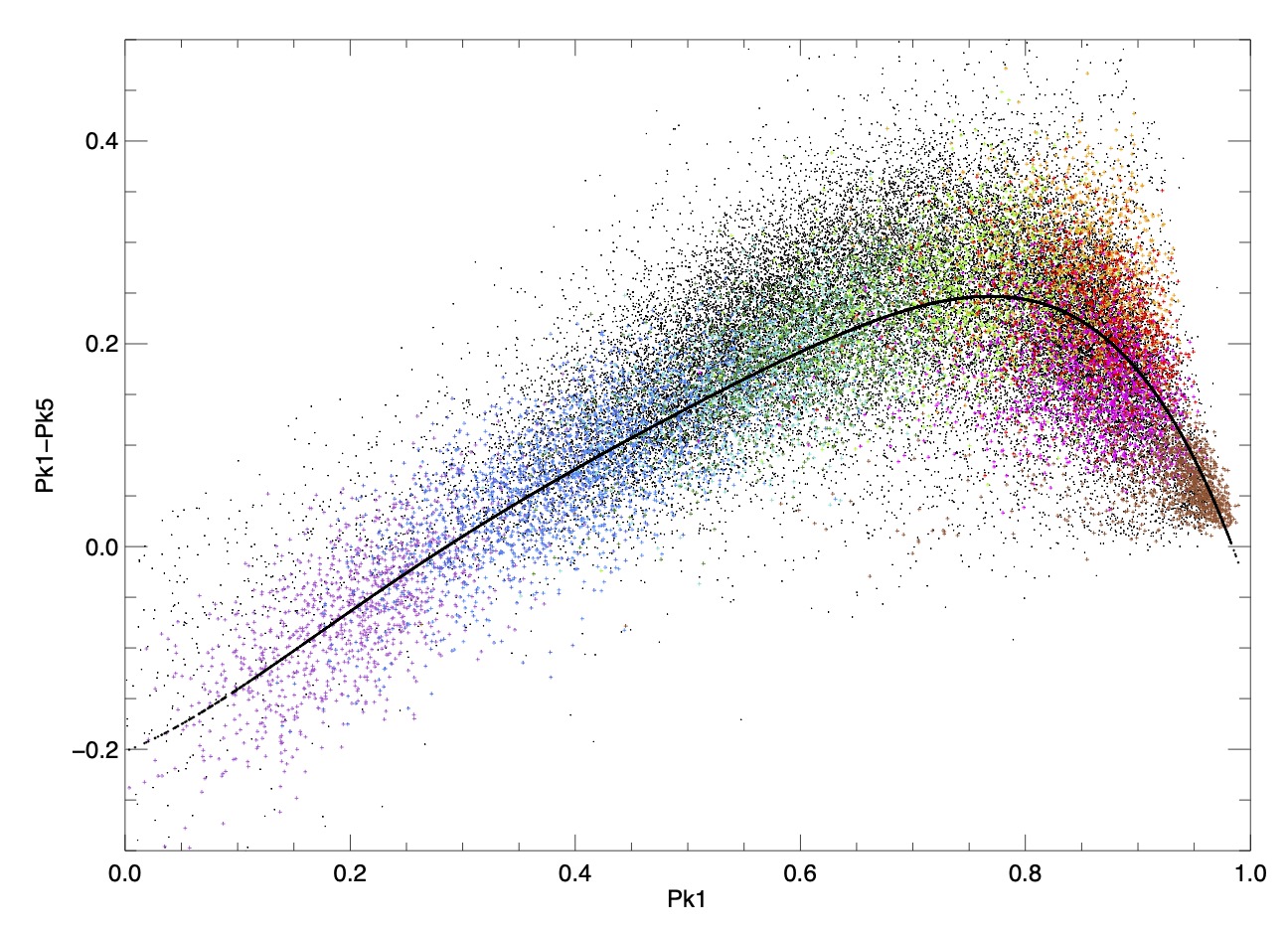}
\end{center}
\caption{The PPP for models with Pk1 compared with the difference between its heights and those of Pk5. Colors stand for lifetimes as in earlier figures. A polynomial fit to the distribution is also shown. It is negative at small values of Pk1 because, as seen in Fig.\ref{fig:combppp} the fit to the observations crosses above the equality line there. }
\label{fig:diffppp}
\end{figure}

Because the Pk1/Pk3 PPP is crowded at the upper right corner and thus has some trouble distinguishing between lifetimes from 10-50 we decided to try the same procedure with another diagnostic. Motivated by the idea that very long lifetimes should have higher correlations between ACF peaks for several rotations (the spot distribution remains quite similar), we tried using the difference Pk1-Pk5 as the ordinate for a PPP.  Figure \ref{fig:diffppp} shows that this produces a differently shaped PPP that offers a little more separation between the points from various long lifetime models. In this case it is more important to compute the shortest distance to the observational fit since the curve turns over and points near the hump overlap in Pk1 but not in Pk1'. The lower panel of Fig.\ref{fig:lifeprob} shows the resulting distributions of the peak difference at different lifetimes. They are relatively close to the ones from the first method (which is somewhat reassuring) but do provide greater separation between the means of the distributions at long lifetimes. In the end we decided to use the fit points from both methods jointly to derive a single polynomial fit to the Pk1 vs log($L_{rot}$) curve.

\subsubsection{Neighborhood Density Distributions \label{sec:PkDens}}

We also tried a third method of producing lifetimes from the Pk1 values. This begins with the same PPP as the first method, but instead of using the histograms of values of Pk1 associated with each lifetime it more directly measures the densities of model points that come from different lifetimes near a specific value of Pk1. The actual procedure is quite simple, although it took some experimentation to optimize the values of its parameters. Given a location in the PPP it takes all the model points for the lifetimes shown in Fig.\ref{fig:combppp}. It counts how many model points lie within a radius of the given location and stores their lifetimes. We ended up using a radius of 0.06 in the PPP. If there are fewer than 10 points within this radius, it does not compute a lifetime. Otherwise, it takes the collection of selected points and finds their average by adding up all their $L_{rot}$  values and dividing by the total count. At the short lifetime end there are relatively few lifetimes represented, and the radius is much smaller than the spread of those points in the PPP. At the long lifetime end there is a much greater mix of points within the radius and their distributions get denser as one moves to the upper right.

We compared the results from the first method with this one. They are fairly consistent, with a 1-sigma difference of about 1.5 in $L_{rot}$. They differ most near short lifetimes where the models scatter most. There the polynomial fit method tends to yield a slightly longer lifetime than the density method as the distribution flattens. This does not mean too much because neither method is very precise at that end, and it is clear that the lifetimes are short for low Pk1 values. 

For the analysis below we elected to use the combined fit from the first two methods. The coefficients of a fifth-order polynomial fit between Pk1 and log($L_{rot}$) are [0.0878629,2.5032105,-14.9314365,51.6520500,-71.0107193,33.9947739]. This polynomial is used to convert the Pk1 values for each Kepler star to a value of log($L_{rot}$) for that star. It should be clear that the inferred values of $L_{rot}$ are uncertain by at least 2-3 rotations simply due to the overlap of model lifetimes at given values of Pk1. As mentioned above there are also unknown systematic uncertainties which could easily be larger. Thus, although we proceed by taking all the inferred stellar values of $L_{rot}$ seriously, the only truly firm conclusions can be drawn from how many stars lie in each of the broad lifetime groups (short, moderate, long) that were defined near the end of \S \ref{sec:PkDistr}.

\section{Results \label{sec:Results}}  %***********************

\subsection{Distribution of Spot Lifetimes with Rotation Period  \label{sec:LifeResults}}

\begin{figure}
\begin{center}
\includegraphics[width=\linewidth]{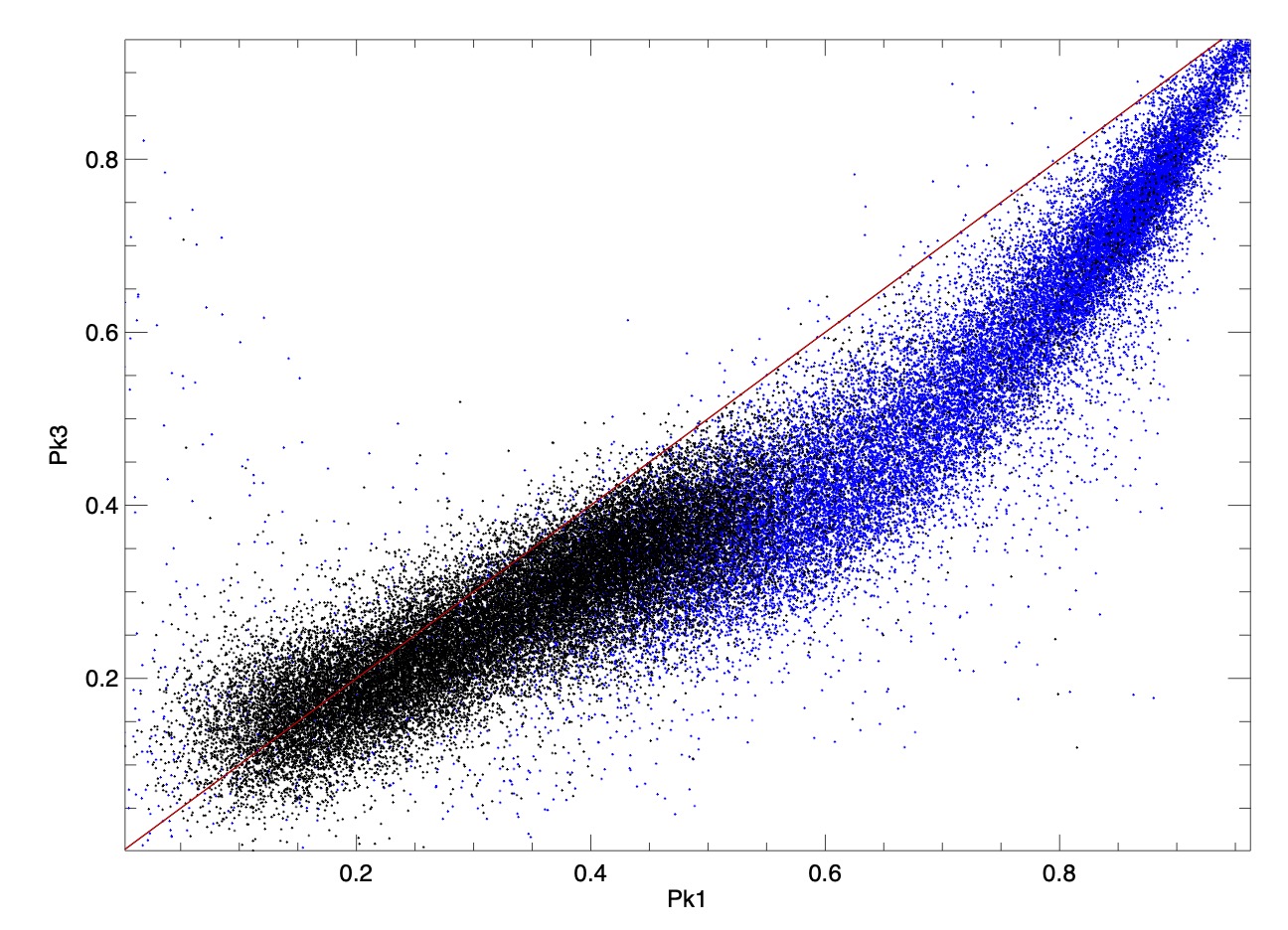}
\end{center}
\caption{The PPP using ACH peak heights for the NP21 stars (black) that MMA14 did not publish periods for, plotted over the MMA14 stars (blue). It is clear that most of the NP21 stars are at the lower left of the PPP meaning they have low autocorrelation strengths that will imply short $L_{rot}$. Those with Pk1 less than 0.2 increasingly have Pk3 a little stronger than Pk1 which is a source of increasingly long pseudo-periods. }
\label{fig:nonMQPks}
\end{figure}

Before considering the results on spot lifetimes we take a digression into the question of what can be done with the stars that don't have currently derived rotation periods -- the NP21 sample discussed in \S \ref{sec:sample}. It is reasonable to suspect that these are the older stars in the Kepler field whose periods are harder to discern. In lieu of a known rotation period we simply used the temporal location of the tallest of the first 3 of their ACF peaks as the anchor for converting to ACH; we refer to these as pseudo-periods. This method also reproduces the vast majority of the MMA14 periods for that sample as we demonstrate below. That is not surprising since to first order it is their method. This enables the analysis for classifying spot lifetimes to be extended to stars without a known rotation period, although one might certainly be more comfortable with the results from ``known" rotation periods. The PPP for the NP21 stars is shown in Fig.\ref{fig:nonMQPks}. It is clear that most of the NP21 stars in Fig.\ref{fig:nonMQPks} will have short derived spot lifetimes because of their low values of Pk1 and Pk3. That result does not depend on what we think their rotation periods are.

\begin{figure}
\begin{center}
\includegraphics[width=\linewidth]{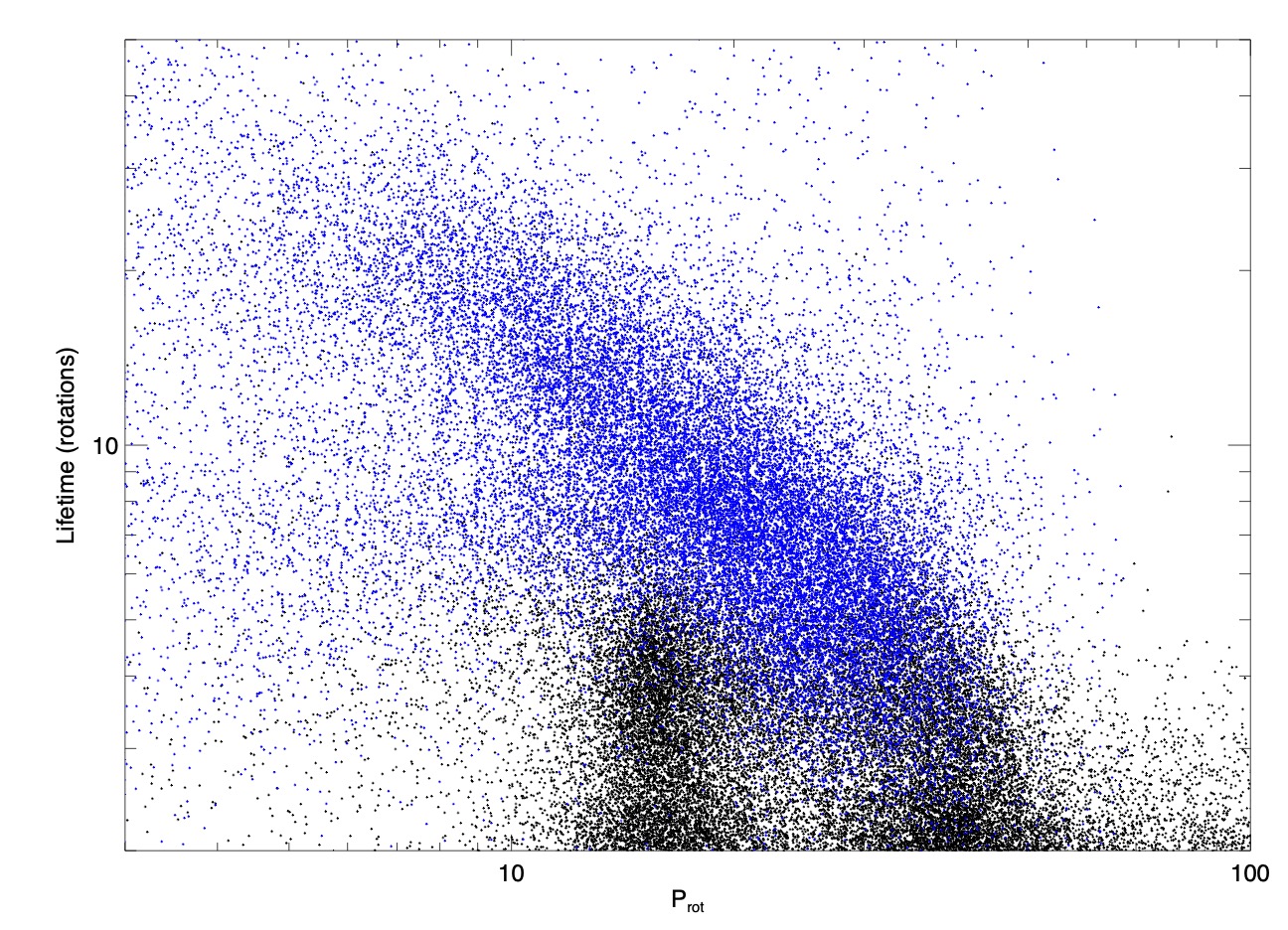}
\end{center}
\caption{The behavior of both the MMA14 sample of Kepler stars with known rotation periods greater than 3 days (blue) and the NP21 sample of main-sequence stars with pseudo-periods (black) as a function of our derived spot lifetimes in units of the stellar rotation periods ($L_{rot}$). Note that both axes are logarithmic. }
\label{fig:RotRel}
\end{figure}

We now discuss the relationships between our derived spot group lifetimes and the stellar rotation periods. We first consider lifetimes in units of the rotation period ($L_{rot}$). Figure \ref{fig:RotRel} shows the result for the MMA14 sample (blue) and for the NP21 sample (black). There is a very clear shape to the MMA14 relation: the stars lie mainly along a trend from long lifetimes at short periods to short lifetimes at long periods. Most of the MMA14 stars have $L_{rot}>2$ rotations and half of them have $L_{rot}<8$. There is a barely discernible diffuse branch of more rapid rotators (periods less than 12 days) with shorter spot lifetimes (less than 8 rotations), and an even vaguer gap in between them and the longer lifetime branch. We separately examined the MS and SG subsets of the MMA14 sample. The SG sample is of course biased to warmer stars that have potentially had time to evolve. The two samples look quite similar, with a slight propensity to shorter lifetimes for MS stars compared with the SG stars.

There is also clump of about 400 MMA14 stars at very short periods (less than 3 days) and very short lifetimes that is not shown in Fig.\ref{fig:RotRel}. Some of these are more likely pulsators than very rapid rotators with starspots. There are few extremely young stars in the Kepler Prime field, and K2 found that light curves from such stars are extremely repeatable and would show up as long spot lifetimes in our method. Furthermore over 70\% of these stars are hotter than 6000K. It is possible that some of them represent a class of stars with very thin convection zones that are barely magnetically braked (hence the rapid rotation) and whose spots are quite short-lived due to the shallow convection. We will return to this topic near the end of this section.

Figure \ref{fig:RotRel} confirms that the NP21 sample is almost entirely composed of stars with short (less than 5 rotations) spot group lifetimes and many are shorter than 2 rotations. There is an unexpected large clump of stars with pseudo-periods between 12 and 24 days. We had seen hints of this phenomenon in earlier unpublished work trying to improve the yield of rotation periods from the Kepler data, and here it is very clear why methods relying on the relative strengths of ACF peaks tend to yield such a result. If these pseudo-periods were correct these stars would violate the relations between range, SDR, and period shown in \citet{Basri18}. Presumably MMA14 did not think they were trustworthy and so did not include them. It is beyond the scope of this paper to find a convincing methodology and justification for doubling pseudo-periods below 24 days (moving that clump to between 24 and 48 days), but that would fix a number of puzzles. In particular, the shorter-period clump of NP21 stars would then lie in the same region as the longer-period clump and fit onto the end of the MMA14 relation. We intend to pursue this question in the near future, but resist making an {\it ad hoc} adjustment here. 

If we presume that the shorter pseudo-periods are misleading, then the NP21 stars are largely older (several Gyr or more) with longer rotation periods. Since the NP21 sample is even larger than the MMA14 sample, this would then be where most of the Kepler solar-type stars lie: stars like the Sun or older whose short $L_{rot}$ and low photometric range makes them difficult to derive a rotation period for. One could assert that most of the NP21 stars have relatively weak (low peak height) ACFs because of the relatively transient nature of their spot groups. That would be a circular argument were it not for the calibration from models (for which it holds). 

It seems reasonably likely from Fig.\ref{fig:nonMQPks} that a small minority of the NP21 stars have light curves from which rotation periods could be confidently derived -- the ones with ACH Pk1 heights greater than about 0.6 that overlap with the MMA14 stars in the PPP. If we presumed that the shorter pseudo-periods were correct then there would be a large new class of stars with relatively rapid rotation (twice solar) that have smaller photometric ranges (almost all below 2 ppt) and short spot lifetimes. At this time we do not think that is the right presumption to make. In principle one could check this by measuring rotation periods using Ca II lines or another activity diagnostic, or by Doppler broadening of spectral lines.

\begin{figure}
\begin{center}
\includegraphics[width=0.7\linewidth]{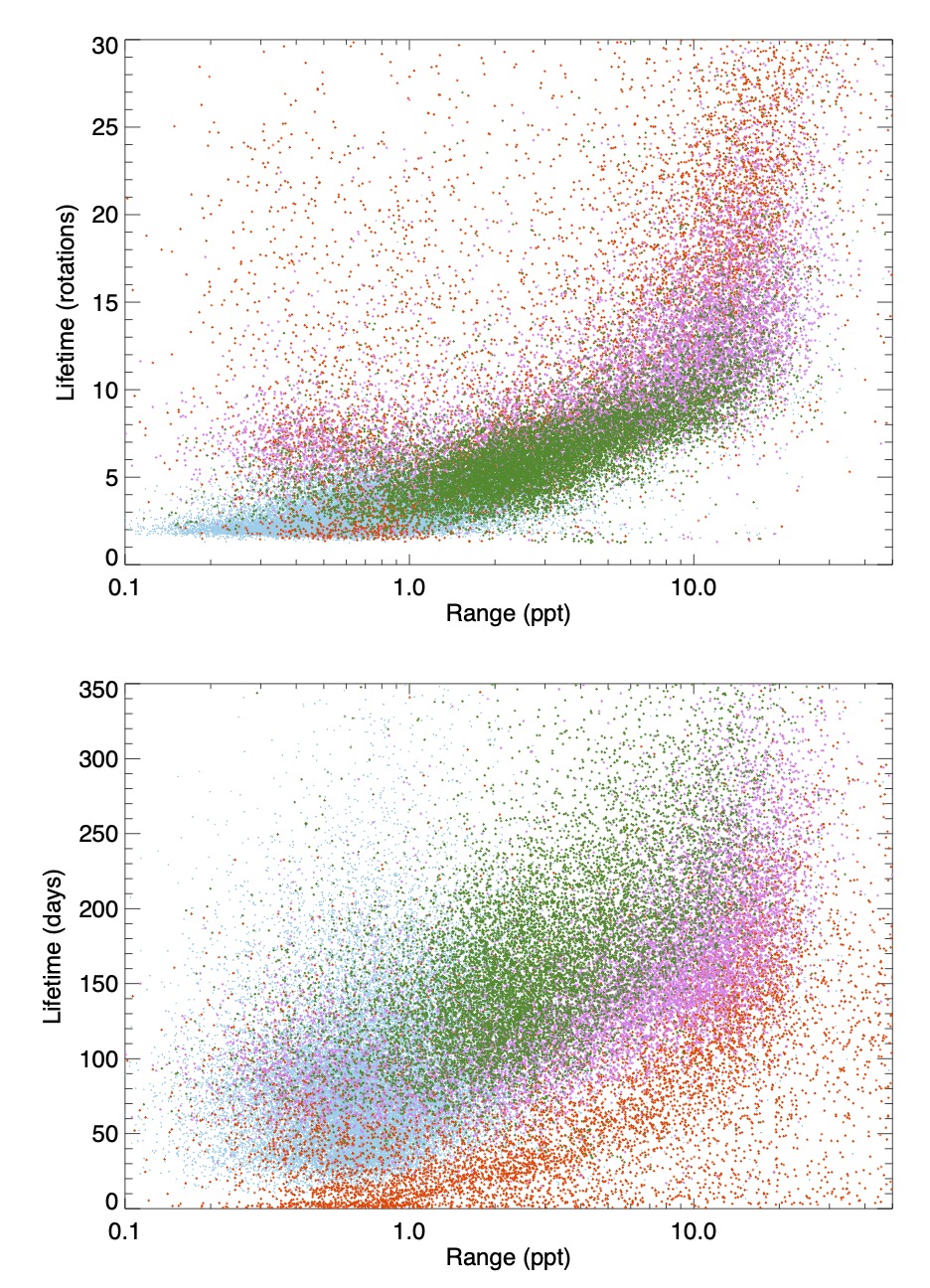}
\end{center}
\caption{The relation between physical spot lifetimes and the photometric range (in parts per thousand) for the full sample. The NP21 sample is in black and the MMA14 sample is split into 3 parts by rotation period. The reddish points have $P_{rot}<10$ days, green points have $10<P_{rot}<20$ days, and light blue points are for $20<P_{rot}<40$ days. The upper panel has $L_{rot}$ and the lower panel has $L_{day}$ on the ordinate. The abscissa is the photometric range (displayed logarithmically). More rapid rotators tend to have longer $L_{rot}$ and shorter $L_{day}$. The NP21 sample is almost all at low ranges.  }
\label{fig:Rangelife}
\end{figure}

One of the conclusions of GCH17 was that stars with larger ranges have longer lifetimes. The upper panel of Figure \ref{fig:Rangelife} shows that is indeed the case for $L_{rot}$. We do not interpret larger range as necessarily implying that there are larger spots as they did, although that is one possibility. The models in BS20 (cf. their Fig. 12) show that the number of spots can also influence the photometric variability, and it is even possible for the range to decrease with longer lifetimes. BS20 also distinguish between range and variability because their models produce absolute intensity variations due to coverage changes. The relevant quantity when comparing to Kepler observations is what they call ``variability" (the median dip depth) but that is nearly equivalent to ``range" for keplerized light curves. The primary factors that increase variability in the BS20 models are the number of spots and higher inclinations. A large variability is primarily diagnostic of an asymmetric spot distribution rather than total spot coverage. The distributions of both variability and coverage get broader with increasing lifetime; the star can find itself with spot distributions more distinct from each other if spots come and go less frequently. BS20 did not directly test the effect of increasing spot size. 

\citet{Basri18} have already presented a detailed examination of the relation between rotation period and photometric range for stars of different temperatures. In the upper panel of Fig.\ref{fig:Rangelife} it is clear that the slower rotators have a fairly well-defined relation between range and $L_{rot}$. The NP21 points (light blue) have both long periods and the lowest ranges. The long period MMA14 points (green) extend the relation to larger ranges and longer spot lifetimes, but not above about $L_{rot}\sim 8$. The moderate and short period MMA14 points lie mostly at the higher ranges and larger values of $L_{rot}$. There is, however, a set of them that lie at the low range end with $L_{rot}$ values between 5 and 10. Some of them also scatter to large values of $L_{rot}$ at all ranges. The small group of 400 short-period short-lifetime hot stars referred to above is visible as the small reddish streak on the bottom of the upper panel between ranges of 0.4 to 1.0.

After the initial submission of this paper we became aware of a very recent paper by \citet{Sant21} that is an extension of the work of GCH17. They attempt to refine the use of the decay of the ACF to infer starspot lifetimes. The signature of a regular decay is that at a given value of Pk1, the Pk2, Pk3, Pk4 and Pk5 values are each lower than the previous peak. It can be inferred from Fig.\ref{fig:McQACF} for observations or Fig.\ref{fig:mod0pkh} for pure models that the ACF only decays in such a regular fashion for Pk1 values above about 0.7. Their paper uses a spot modeling protocol that shares many similarities with ours. One difference is that they don't use fixed lifetimes but rather random spot areas (with a maximum limit) coupled with the Gnevyshev–Waldmeier rule to set the lifetimes. Both the light curves and ACFs shown in their paper are very ordered however, unlike most of the Kepler data and all but the longest of our model light curves, although there is not enough information for us to know how typical they are. It is only so organized that basing a lifetime metric on the ``decay" of the ACF seems to us likely to work. 

We looked at the statistics of how many of the stars and models in this paper exhibit descending ACH peaks for the first few peaks that would be amenable to a decaying ACF analysis. For our models, only 4\% do so at $L_{rot}=2$, about half at $L_{rot}=5$, 80\% at $L_{rot}=10$ and nearly all above that. In the MMA14 sample 46\% do (71\% of those also have Pk1 greater than 0.7) while in the NP21 sample only 9\% do.  In any case \citet{Sant21} do not attempt to derive the lifetimes exhibited by the bulk of the Kepler stars as is the main thrust of this paper, so we cannot do a detailed comparison between their results and ours.

\subsection{Relation to Stellar Parameters  \label{sec:starpar}}

Because we know the rotation periods in days of all the MMA14 stars and have pseudo-periods for the NP21 stars we can convert their rotational lifetimes ($L_{rot}$, derived from fitting with models) to physical spot group lifetimes $L_{day}$. The lower panel of Fig.\ref{fig:Rangelife} has a somewhat different appearance from the upper panel because of the influence of the actual rotation period on the lifetime in periods. These physical spot group lifetimes are more spread out in days, and the ordering of the short to long period stars is somewhat reversed. Now the short period stars lie at the bottom of $L_{day}$ values at all ranges. This is just because the conversion of $L_{rot}$ to $L_{day}$ means that a short $P_{rot}$ star with long $L_{rot}$ can end up having a relatively short $L_{day}$. For example, a star with $P_{rot} = 5$ and $L_{rot} = 10$  will have $L_{day} = 50$, while a star with $P_{rot} = 25$ and $L_{rot} = 2$ has the same $L_{day}$. The moderate and long period stars are more cleanly sorted to longer lifetimes, again more independently of range. The visible part of the NP21 sample is heavily influenced by the numerous moderate pseudo-period group (that we are not sure about). Some of the longer pseudo-period points from NP21 are hidden by the MMA14 points that are plotted over them and some are visible in the upper scattered points.

\begin{figure}
\begin{center}
\includegraphics[width=\linewidth]{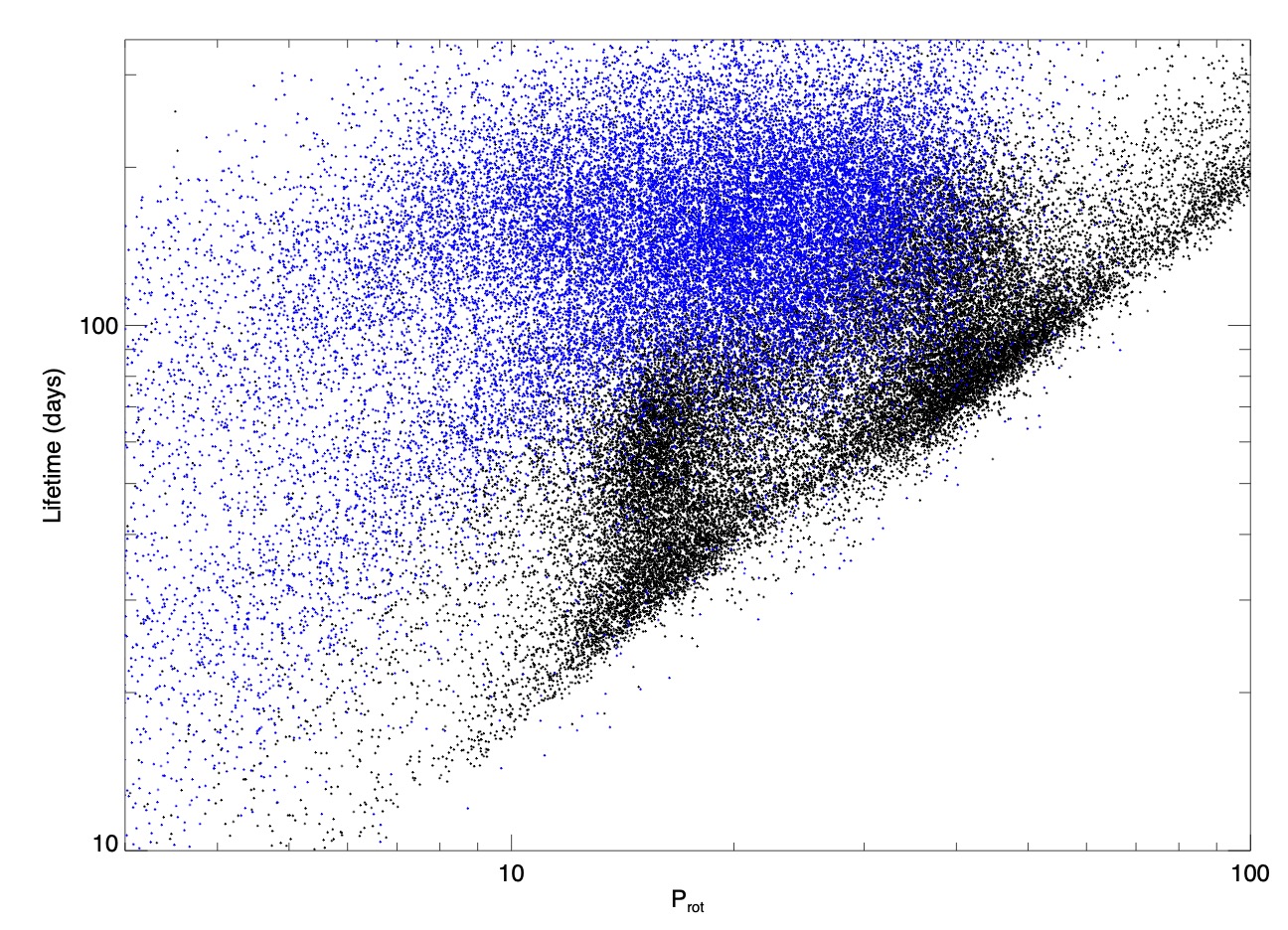}
\end{center}
\caption{Like Fig.\ref{fig:RotRel} except with the physical lifetimes $L_{day}$ as the ordinate. The bottom boundary of points follows the $L_{day}\leq P_{rot}$ line for lifetimes of one rotation or less. }
\label{fig:DayRel}
\end{figure}

Figure \ref{fig:DayRel} shows the behavior of the stars in both samples for $L_{day}$ as a function of $P_{rot}$ (both in days). The MMA14 sample is again blue and the NP21 sample is black. The same qualitative features as seen in Fig.\ref{fig:RotRel} are still visible, but in different places with different appearances. It is more obvious that the pseudo-period NP21 stars contain two groups, one near the low lifetime boundary and clumps just under the MMA14 stars with moderate to long periods. If the pseudo-periods in the shorter-period group are half of what they should be because the half harmonic has a slightly stronger autocorrelation than the true period for these poorly autocorrelated stars then they would neatly join the longer pseudo-period NP21 stars and continue the MMA14 distribution down to the low lifetime boundary. The MMA14 sample appears to have a nearly constant but widely spread set of $L_{day}$ values between about 50 to 250 days, but the two vague branches at periods below about 12 days mentioned above are visible, now pointing down to the left at lower and higher lifetimes with a vague gap between them. The SG subset is shifted on average to $L_{day}$ values about 20 days less with a little excess at low values due to the bias toward hotter stars.

\begin{figure}
\begin{center}
\includegraphics[width=\linewidth]{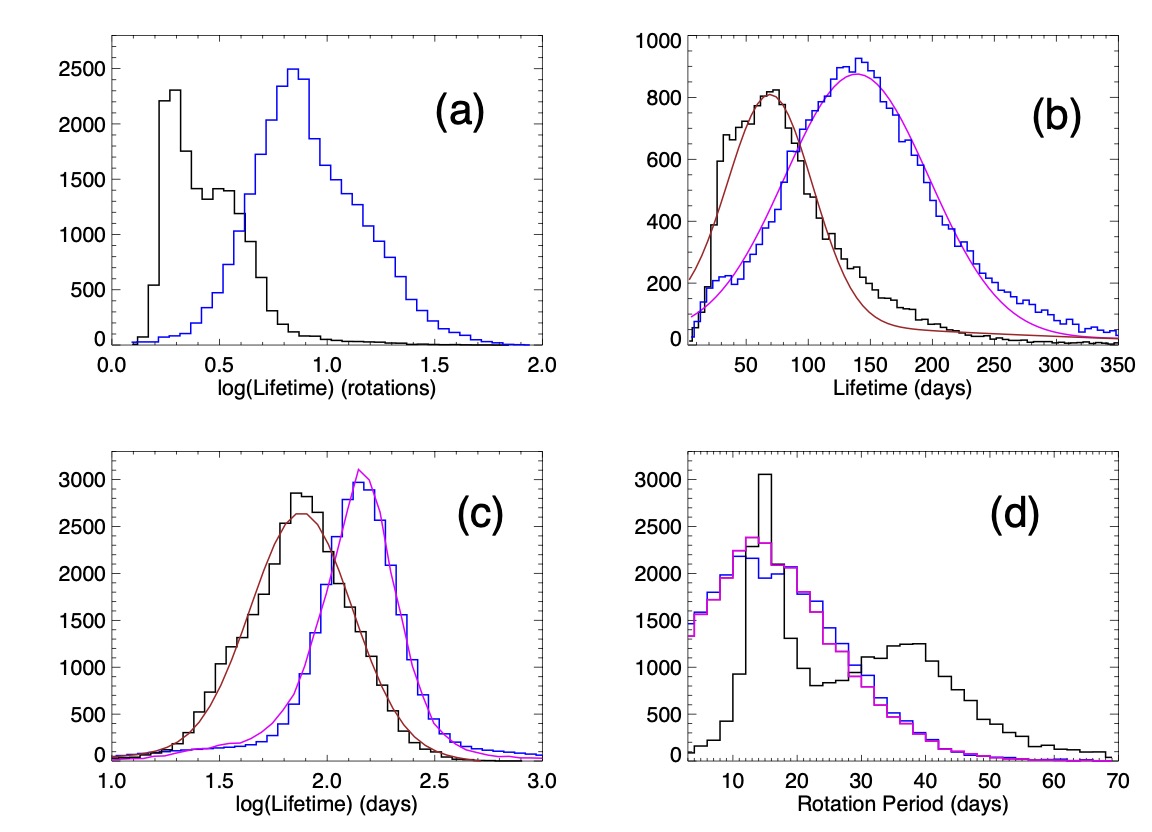}
\end{center}
\caption{ Histograms of interest. a) The logarithmic distribution of log($L_{rot}$). The blue curve is for the MMA14 sample and the black is for the NP21 sample. The latter is divided by 2 to fit on the same scale. b) The distributions of $L_{day}$ for the same two samples. The orange curve is a Gaussian fit to NP21, and the magenta curve is a Gaussian fit to MMA14. c) The distributions of log($L_{day}$) with the same representations as (b). d) The distributions of the MMA14 rotation periods (blue), the periods for that sample derived using the highest of the first 3 ACF peaks (magenta), and the periods for the NP21 sample derived the same way (black). }
\label{fig:lifehists}
\end{figure}

We now examine the general distributions of $L_{rot}$ and $L_{day}$ independent of rotation period. The top panels of Fig.\ref{fig:lifehists} show the overall distributions for both lifetime units for stars of all temperatures and rotation periods. The distribution of log($L_{rot}$) is not Gaussian (panel a), displaying a distinct ``knee" on the long lifetime side. Both the NP21 and MMA14 samples rise faster than they fall as period increases, peaking at 2 and 7 rotations respectively. The distributions of $L_{day}$ are more Gaussian (panel b) with a little excess power at long lifetimes, peaking at 80 and 140 days respectively. Both distributions of log($L_{day}$) are quite close to Gaussian, meaning that $L_{day}$ has a lognormal distribution (panel c) for both samples. This would result if physical spot group lifetimes are caused by at least 4 or 5 independent parameters, each of which are distributed normally. It is interesting that the NP21 and MMA14 samples are separately lognormal. 

Sunspots display a lognormal distribution in their sizes (thanks to Dr. Solanki for bringing this to our attention) so it is not surprising that their lifetimes show such a distribution as well \citep{Baum05}, with some complications for small spots. It is not straightforward to compare our distributions to the solar case. The NP21 linear peak in $L_{day}$ is at about 3 solar rotations, however, which is definitely longer than for the Sun. Of course it is composed of stars of many effective temperatures and gravities. They undoubtedly have different typical spot sizes and lifetimes, so it is not hard to imagine that there are indeed several normally distributed parameters that go into making up the final distribution. It is not surprising that the most common lifetime in the MMA14 sample is substantially longer than for sunspots since that sample is biased towards stars that are faster rotators (younger) than the Sun. It will be possible in the future to use our methodology to break the sample up into smaller and more focused sub-samples to investigate the behavior of spot lifetimes in a more systematic way; we begin that process below.

Panel (d) of Fig.\ref{fig:lifehists} shows the distributions of the periods and pseudo-periods we are using for the stars in the two samples. The black histogram has the published periods from MMA14 and the green histogram has the pseudo-periods we derive from the same sample. The two are remarkably close, indeed the only real difference is near 15 days where the MMA14 sample has a small dip while the pseudo-periods show what looks like a more normal distribution that peaks in about the same place. This may be related somehow to the very odd and strong peak of periods in that same region in the NP21 sample we have mentioned several times before. Because this paper is not primarily focused on the determination of rotation periods using autocorrelation functions we leave this mystery to be explained in future work. Otherwise the NP21 sample peaks at around 40 days, which might be expected from gyrochronology arguments, and continues at a low level out beyond 100 days. We also don't have much confidence in those very long periods; they are unlikely to be detectable in Kepler data and simply reflect the fact that when the light curve is quite aperiodic, the low level autocorrelation peaks are not organized by decreasing heights at smaller shifts and sometimes the highest peak can occur quite far out.

\begin{figure}
\begin{center}
\includegraphics[width=0.8\linewidth]{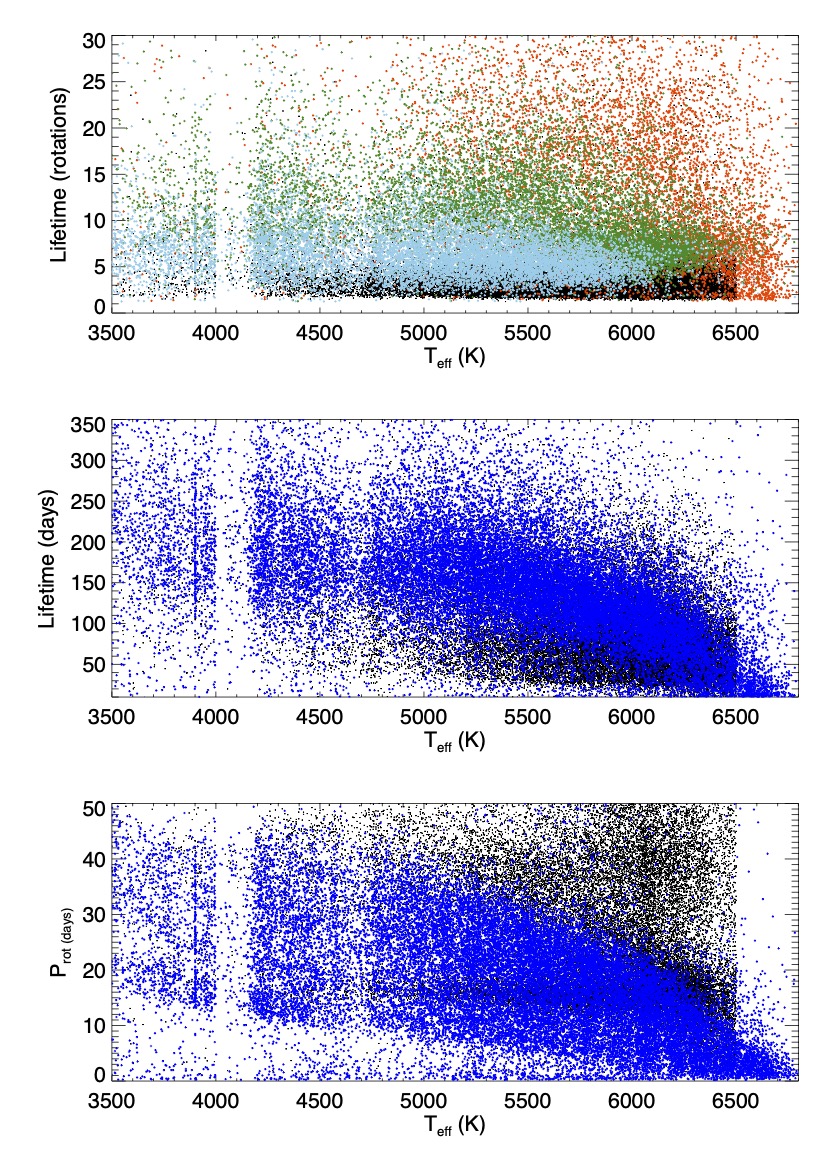}
\end{center}
\caption{Top panel: The distributions of the spot lifetimes for the MMA14 and NP21 (colors as in Fig.\ref{fig:Rangelife}) Kepler stars as a function of effective temperature. Middle panel: the distributions for $L_{day}$ in the full MMA14 sample (black) and NP21 sample (light blue) as a function of effective temperature. Bottom panel: the distributions of $P_{rot}$ for the two samples (as in the middle panel).  }
\label{fig:TeffLife}
\end{figure}

Because both samples have stellar parameters from Gaia we can also examine the overall behavior of spot group lifetimes with effective temperature and radius. The top panel of Fig.\ref{fig:TeffLife} shows that $L_{rot}$ does not depend very much on temperature; there is a broad range of a particular value of $L_{rot}$ over many temperatures. The concentration of stars at solar temperatures is determined mostly by the composition of the sample rather than a dependence of $L_{rot}$ on temperature. The dependence of $P_{rot}$ on temperature is visible in the color segregation and the bottom panel shows that more directly. The middle panel of Fig.\ref{fig:TeffLife} shows the result for $L_{day}$ for the whole sample. We examined this relation for the MS and SG stars separately in the MMA14 sample; we found essentially the same distributions so apparently stellar gravity is not an important variable (at least in the range we are testing).

The middle panel of Fig.\ref{fig:TeffLife} shows a tendency for the physical lifetime to steadily increase as stars get cooler, along with a spread at each temperature that gets larger for cooler stars. The first of these is the same result that GCH17 found. We argue that this trend with temperature is primarily due to the fact that cooler stars tend to have longer rotation periods in this sample. In fact, a plot of temperature vs. rotation period (bottom panel of Fig.\ref{fig:TeffLife}) looks rather similar to the middle panel. It is not yet clear to what extent this is a result of a bias introduced by the fact that cooler stars tend to have larger ranges (\cite{Basri11}) and so are easier to detect rotation periods for. It is also important to note that there are many M dwarfs that have short rotation periods (less than 10 days) which are significantly underrepresented in the MMA14 sample. TESS is supplying many new light curves from that population but unfortunately they mostly do not extend as long as desirable for a project like this.

\begin{figure}
\begin{center}
\includegraphics[width=\linewidth]{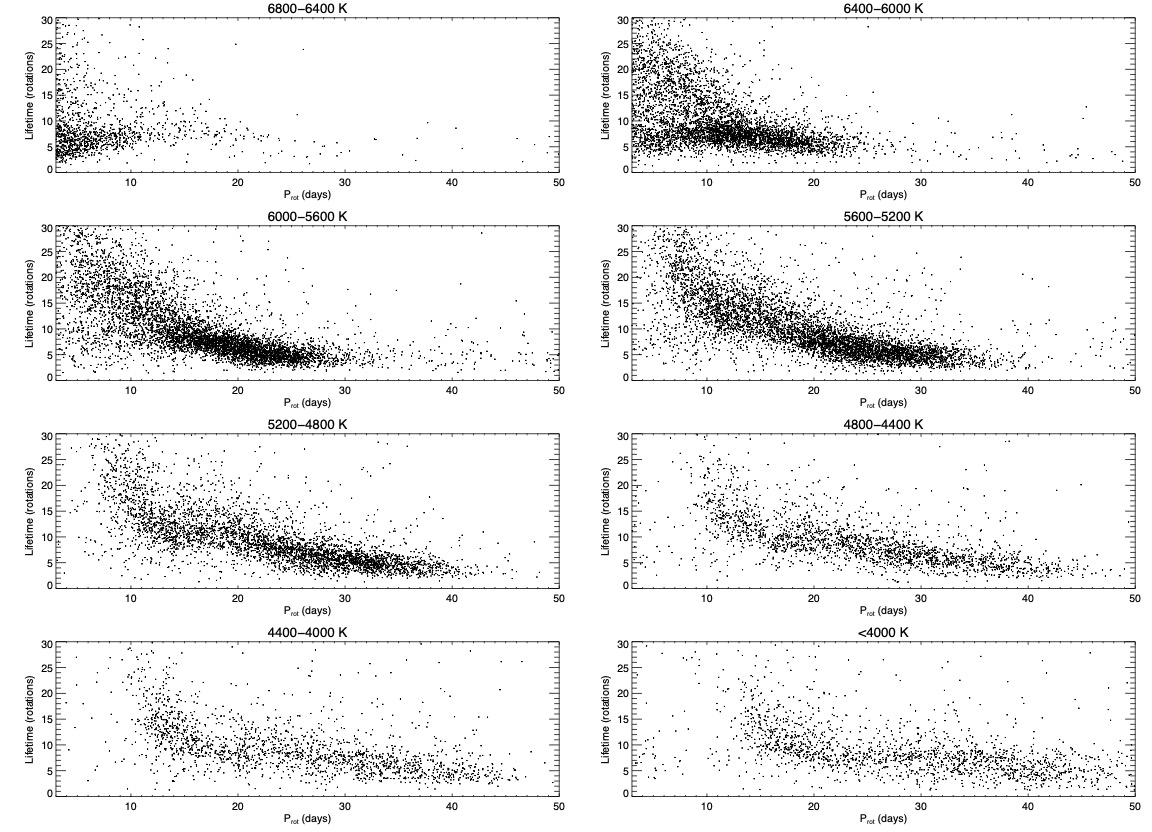}
\end{center}
\caption{ A separation of the relations for the MMA14 stars between $L_{rot}$ and rotation period into 8 groups of effective temperature. Except for the top two panels their appearance is qualitatively similar except for the prevalence of different rotation periods at different temperatures (longer periods for cooler stars). }
\label{fig:TempLRot}
\end{figure}

\begin{figure}
\begin{center}
\includegraphics[width=\linewidth]{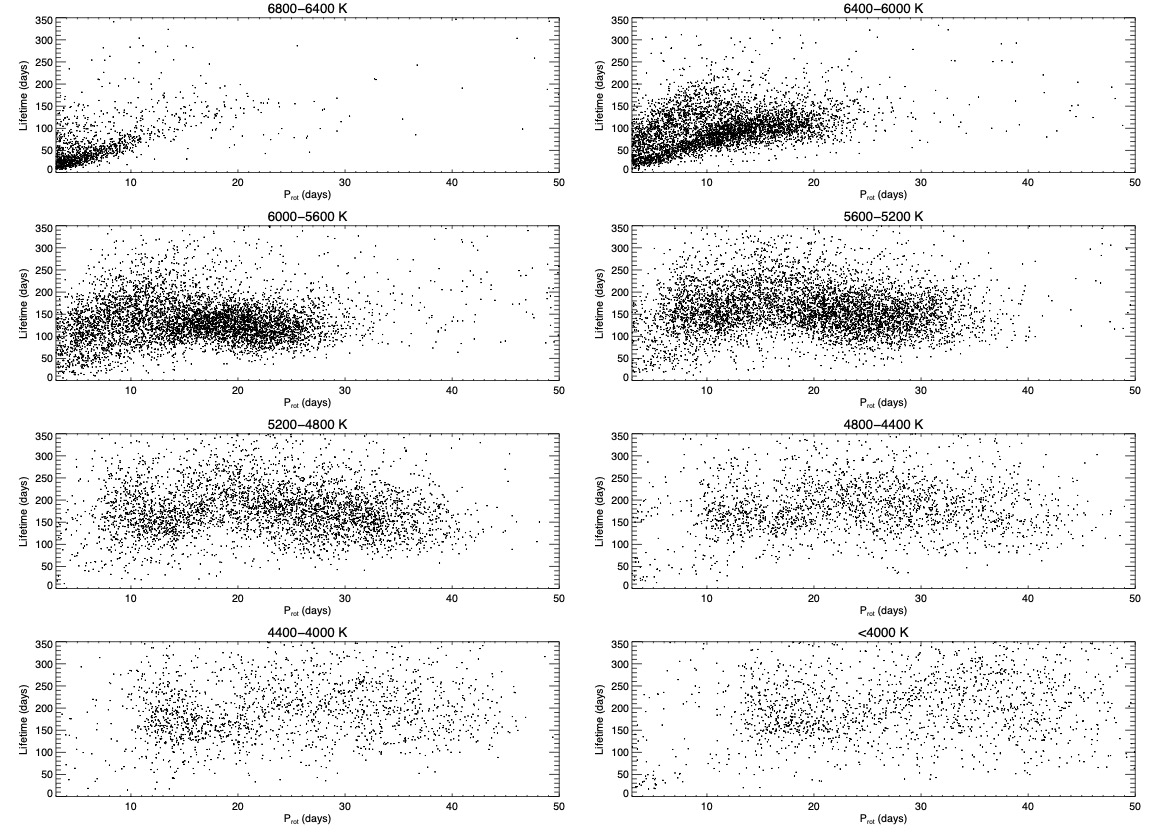}
\end{center}
\caption{ A separation of the relations for the MMA14 stars between $L_{day}$ and rotation period into the same 8 groups of effective temperature as the previous figure. Stars above 6000K tend to have shorter physical spot lifetimes and this group disappears cooler than about 5600K. Below that temperature the lifetimes occupy the same region of values but the scatter in each panel becomes larger as the stars become cooler. The longest physical lifetimes are associated with the cooler, slowly rotating stars. }
\label{fig:TempLday}
\end{figure}

In Figs. \ref{fig:TempLRot} and \ref{fig:TempLday} we continue discussing the relations between lifetime and rotation period but now break the stars into 8 temperature groups. The solar-type stars constitute the majority of the MMA14 and NP21 samples (by design) so they are primarily responsible for the general appearance of Figs. \ref{fig:RotRel}-\ref{fig:DayRel}. The hottest stars lie in the short-period short-lifetime corner for both $L_{day}$ and $L_{rot}$. The prevalence of high values ($>10$ rotations) of $L_{rot}$ for low values ($<10$ days) of $P_{rot}$ increases dramatically below 6400K, then stars that rotate that rapidly become rarer below 5600K. Most of the stars in the MMA14 sample have $L_{rot}<10$ below 6000K and their population density shifts to larger values of $P_{rot}$ as the temperature decreases. The shape of the relations in Fig.\ref{fig:TempLRot} doesn't change much with temperature below about 6000K other than the density shift, although a larger percentage of cooler stars with periods greater than 20 days rise above the $L_{rot}<10$ main branch (this may not be obvious in the figure because of the lower numbers of cool stars). 

A subtle effect at around periods of 12-18 days is visible in Fig.\ref{fig:TempLRot} (if you look for it) at temperatures between 5200K-4000K, namely a dearth of stars in that period range compared to periods longer and shorter for $L_{rot}>15$. Other (often small) effects in this period range have been noted before. In addition to the dip in the MMA14 $P_{rot}$ distribution seen in the bottom panel of Fig.\ref{fig:lifehists}, this is approximately where there might be a transition from spot to facular domination of light curves (\citet{Mont17}, \citet{Rein19}) and where a kink in the relation between photometric range and rotation relations is seen in stars of these temperatures \citep{Basri18}. It will be interesting to highlight and pursue the various (rather subtle) anomalies that seem to occur in this period range.

In Fig.\ref{fig:TempLday} the short-period short-lifetime branch is very obvious above 6000K, and it grows more diffuse at solar temperatures (6000K-5600K) while the longer period clump at periods of 15-30 days and $80<L_{day}<200$ becomes apparent. The same subtle effect mentioned in the last paragraph can be seen for $L_{day}>200$ as well. It is partly responsible for the vague impression of a clump of cool stars between 10-20 days (moving longward with decreasing temperature) for stars cooler than 5200K. There is a general trend for $L_{day}$ to become longer at temperatures below 4800K and periods greater than 20 days. These cooler stars are responsible for the increasing spread in the lifetimes at longer periods, especially below 5200K. The fact that these stars tend to have larger photometric ranges at a given period is consistent with an inference that some of them have longer spot group lifetimes. In the other direction, longer spot lifetimes could be part of the reason that they tend to exhibit larger photometric ranges. Neither that causality nor its direction has been established at this point. 

We now take a detour to the interesting small set of Kepler stars studied by \citet{Rein20}  (we thank Dr. Reinhold for providing all their KICs). They have similar rotation periods, metallicities, and temperatures as the Sun but larger photometric variability. The Sun is in the third temperature group in Figs. \ref{fig:TempLRot} and \ref{fig:TempLday} at a rotation period of 27 days. It has a very short $L_{rot}$ and $L_{day}$ of less than about 60 days so it sits at the bottom of the relevant MMA14 distributions (more in the middle of the NP21 sample as seen in the middle panel of Fig.\ref{fig:TeffLife}). It has Pk1 heights ranging from 0.15-0.45 depending on which segment of its light curve is used; its median Pk1 is around 0.35 so its $L_{rot}$ is short according to our method (and in reality). Its period is not usually derivable from its light curve when the methodology of MMA14 is applied but of course is known to be about 27 days; its $L_{rot}$ is at most about 2 implying $L_{day}$ of 60 or fewer days. The stars \citet{Rein20} identified as more active than the Sun at the same rotation rate show Pk1 heights between 0.4-0.7 yielding $L_{rot}$ in the range 2-5 rotations. These are in the bulk of solar-type stars in the MMA14 sample but larger than the NP21 stars. 

There are far more solar-type stars that resemble the Sun in our lifetime diagrams for the combined samples than resemble these more variable solar-type stars. The comparison non-periodic sample in \citet{Rein20} also has lifetimes in the 1-3 range, very consistent with the NP21 sample. Their paper's title implies that the Sun is anomalously quiet, but that does not seem true in light of our results, which instead imply that the Sun IS an average solar-type star (reinforcing the arguments in \citet{Basri13}). What \citet{Rein20} have identified is a small set of stars with Sun-like stellar parameters that exhibit larger and more regular photometric variability, perhaps related to their longer-lived spot groups. This is closer to the actual conclusions in their paper, which are that either there is a group of Sun-like stars that are typically more active than the Sun or the Sun exhibits exceptionally greater activity a small fraction of the time and we haven't seen it so far. We applied the filters from their paper (except for metallicity) to our sample and found another 500 stars smaller than 1.3 solar radii that are in the same temperature and rotation period bands but with even longer spot lifetimes than the sample of 369 they found. It will be interesting to pursue the question of what makes these stars anomalously active. 

Following this thread, it is also apparent that among the cool stars there is an even larger percentage of stars with long rotation periods and long spot lifetimes. This implies they have relatively periodic light curves and we know they have larger ranges. We examined a few of these individual light curves and they are indeed both high-amplitude and clearly long-period. To understand this trend better we chose an upper boundary to the bulk of stars in Fig.\ref{fig:TempLday} of $L_{day}\sim 200$. We then looked at the ratio of stars above and below that threshold for all rotation periods as a function of temperature. This is a steadily decreasing function of increasing temperature, well fit by a quadratic between the central temperature of each bin and the ratio of stars above the threshold to those below it. The ratios vary from .025 at 6600K to .114 at 5800K to .601 at 4600K to 1.108 at 3800K. The prevalence of ``anomalous" activity is a reflection of the fraction of stars with these longer lifetimes and larger ranges. The trend and spread to longer spot lifetimes for cooler stars appears to involve something that is probably related to the depth of the convection zone, with spot lifetimes becoming increasingly weighted towards longer values as convection deepens (although shorter lifetimes remain present at some level). This may be helpful to the question of why cooler stars are generally more photometrically variable, and possibly tied to their tendency to have higher total magnetic fluxes.

\subsection{Effects of Inclination and Differential Rotation  \label{sec:indiffrot}}

\begin{figure}
\begin{center}
\includegraphics[width=\linewidth]{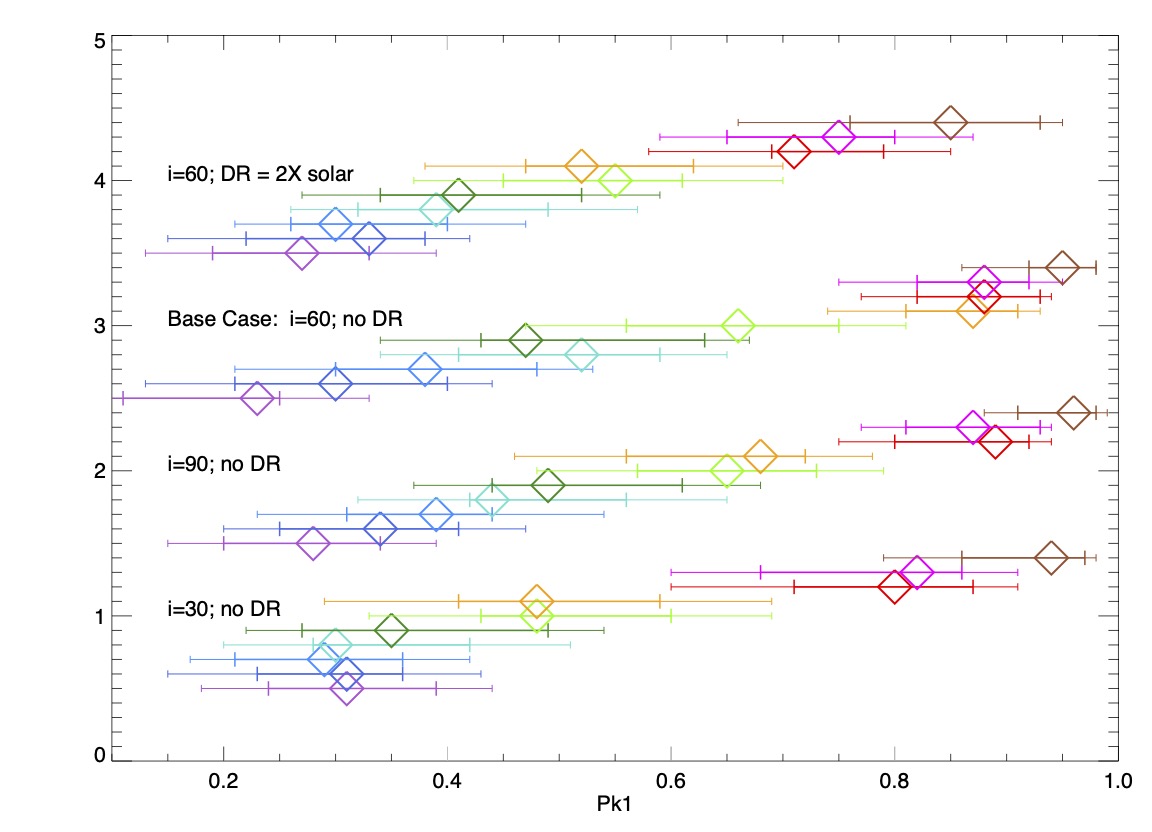}
\end{center}
\caption{ The effects of stellar inclination and differential rotation on the distribution of Pk1 values by lifetime. This is another way of displaying the information in Fig.\ref{fig:lifeprob}. The color coding is the same, and four cases are shown separated vertically (the ordinate is only for placement). Each lifetime generates a distribution of Pk1 values (the abscissa) and the location of each distribution's maximum is shown with a diamond. The half-height extent of the distribution is shown with the thicker line and larger ticks, and the 20\% heights extent by the thinner lines and smaller ticks. The base case in Fig.\ref{fig:lifeprob} is second from the top. The bottom two sets are for the same case, but viewed at stellar inclinations of 90 and 30 degrees. The top case is the base case but with differential rotation of twice the solar shear imposed. } 
\label{fig:HistDist}
\end{figure}

In this section we examine the effects of relaxing some of the assumptions we made for the base case used above. The three we study are: what is the effect of stellar inclination on the inferred lifetimes using the procedures above, what is the effect of differential rotation, and how many periods does a light curve need to contain to provide a reasonable spot lifetime? \citet{Sant21} also looked at these questions. They found that inclination had little effect on their analysis. Our modeling procedure allows us to view exactly the same case from different inclinations, so we can construct PPPs for them as in Fig.\ref{fig:combppp} and then repeat the analysis that led our basic method for inferring lifetimes from a PPP. In Fig.\ref{fig:HistDist} we show the results from the 3 tests we conducted. The case shown in Fig.\ref{fig:lifeprob} is our base case and we characterize the main properties of the distributions in that figure by marking the location of the maximum of the distribution of Pk1 values for each lifetime and the extent of the distribution at 50\% and 20\% of that maximum value in the set of points around ordinate values of 3. The ordinal value of each point has no meaning beyond serving to separate them for clarity. 

The set of points around ordinate 2 are the distributions for the same case viewed with stellar inclination 90 instead of 60 degrees. They are qualitatively the same as the base case, with some re-positioning of a couple of lifetimes between 5 and 12.5 that are closer what one might have expected. The lowest set around ordinate 1 are for inclination 30 degrees and they show more significant changes. The lower lifetimes now all cluster around low values of Pk1, the distributions for $L_{rot}=10,12.5$ are at substantially lower values of Pk1 (just under 5 instead of 6.5 to 7) and $L_{rot}=15,20$ are also lower although $L_{rot}=50$ is not. The low stellar inclination means that many of the spots are visible through more of a rotation, so perhaps it is harder to cause significant period patterns to appear in the light curves. The net effect is a pinching of the Pk1 distribution towards the middle of possible values.

We also imposed a solar differential rotation (DR) law with twice the solar shear on the spots in the base case (starting with the same configuration). Unlike on the Sun, our spots are uniformly distributed over latitude so they do a better job of sampling the full shear. Thus this case is probably at the extreme of how much DR might be observed. The results are shown in the top set around ordinate 4 in Fig.\ref{fig:HistDist}. The Pk1 value for lifetime 2 is moved up closer to that for 3 and 4, but all the lifetimes 5 or longer have reduced values of Pk1. This is not an unexpected result since the differential rotation disturbs the patterns that otherwise exist so they last less long. \citet{Sant21} obtained a similar result. For some reason, the distribution for $L_{rot}=12.5$ (orange) moves around the most in our test cases.

\begin{figure}
\begin{center}
\includegraphics[width=\linewidth]{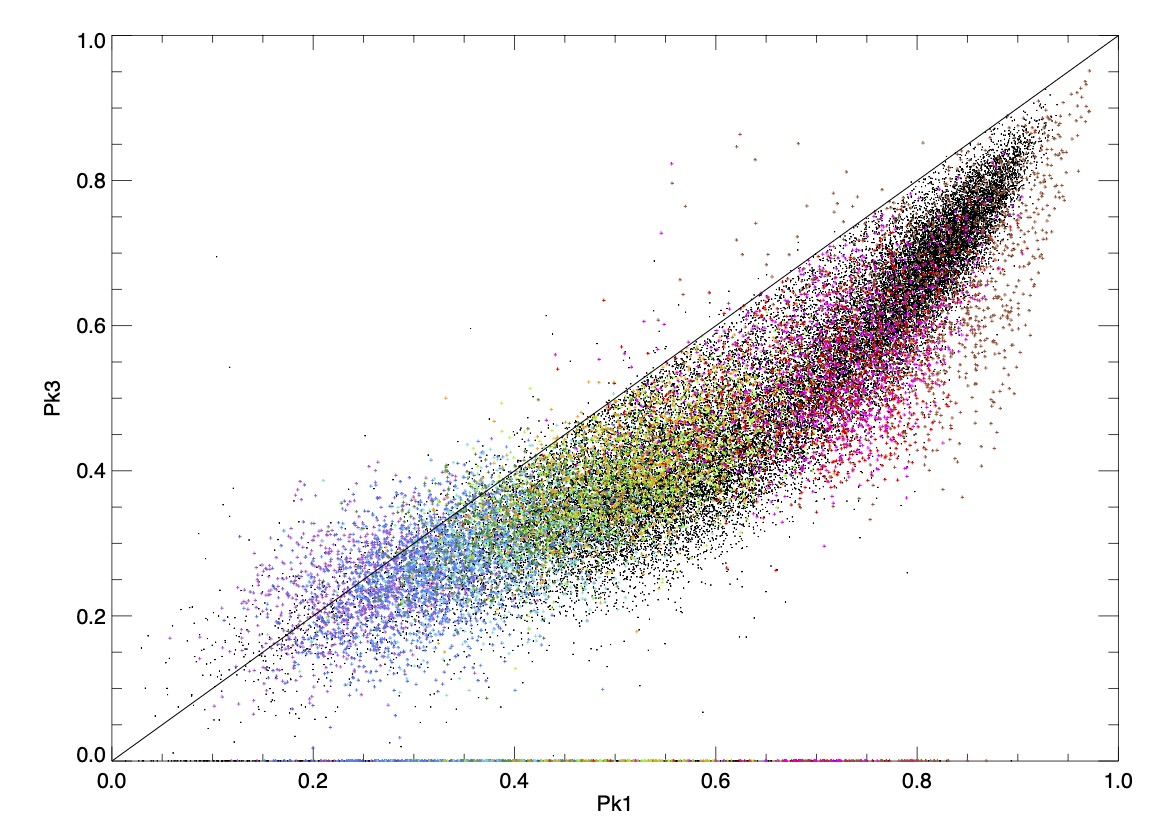}
\end{center}
\caption{ The effect of differential rotation on the distribution of model points from Fig.\ref{fig:combppp}. The shortest lifetimes have Pk1 values a bit higher, but most lifetimes are moved to smaller values of Pk1. The values of Pk3 are more dispersed, especially at longer lifetimes. The model points no longer closely resemble the observations, and the observed points with Pk1 greater than 0.8 are barely reproduced at all. }
\label{fig:Diffrot}
\end{figure}

Fig.\ref{fig:Diffrot} shows that not only are the Pk1 locations at a given lifetime affected but naturally the higher peaks are too. In particular, the Pk1/Pk3 PPP looks noticeably different with and without strong DR (compare with Fig.\ref{fig:combppp}). Of note is the behavior at the long lifetime/high Pk1 end of the PPP. The points for lifetimes longer than about 10 are spread out in Pk3, producing a wider distribution than observed. Their retreat from the highest values of Pk1 also means that the upper end of the Kepler points are no longer covered at all. We tried different mixes of short and long lifetimes, but could no longer find a mix that works to reproduce the observations when strong DR is present. The implication is that the observed stars with long lifetimes are not subject to strong DR. It is possible that the few Kepler points that lie in a wider distribution than the bulk of the points in the upper half of the PPP are examples of stronger DR, but that is certainly not definitive and will require further investigation.

The real question is to what extent these sensitivities affect the estimation of starspot group lifetimes using the methodology of this paper. Returning to Fig.\ref{fig:HistDist} one can look along vertical slices to see which lifetimes are represented in the various cases. It is again clear that the most reliable conclusion that can be made is that there are three basic groups of lifetimes that seem fairly robust against the tests we made. Short lifetimes (1-4) are implied when Pk1 is less than about 0.4, long lifetimes (15 or greater) are implied when Pk1 is greater than about 0.7, and in between are the intermediate cases. Beyond that there is a reasonably robust ordering of lifetime with Pk1, but it does depend somewhat on stellar inclination and differential rotation, and the uncertainties in translating Pk1 to lifetime mean an uncertainty of at least 2-3 and sometime as much as 5 in the inferred $L_{rot}$. We repeat here that there are also other unknown systematic uncertainties associated with both the methods of modeling starspots and the definitions of ``lifetime". 

We also took a brief foray into the question of how long a light curve needs to be (in units of rotation period) to render our procedure effective, as did \citet{Sant21}. We tested sub-segments of model light curves to see how consistent the ACH peak strengths are, by selecting three segments of 25 rotation periods each (at the beginning, middle, and end of each 50 rotation model run). The Pk1/Pk3 PPPs for the segments were then compared against what is found from the full light curves. We found that they look essentially the same but with a little greater dispersion on both axes up to $L_{rot}=12.5$. For $L_{rot}\geq 15$ the PPP also spreads a bit but more importantly develops a tail of points towards lower values of Pk1. This means that one would be more likely to interpret points as having shorter lifetimes than they really did for the longest lifetimes. That is a similar effect to the addition of differential rotation. Presumably things get even more muddled the fewer rotation periods there are in a light curve. 

For the Kepler Mission a star with the solar rotation period and full coverage over 4 years has nearly 50 periods in its light curve, so that is certainly adequate. In any case, stars with long rotation periods have short spot lifetimes so the lack of many periods is less important; one will infer short lifetimes anyway. The stars with long spot lifetimes tend to have shorter rotation periods, so one needs a shorter length of observations to cover the requisite number of periods. This means that analyzing the Kepler Prime light curves is fairly safe, but there would be questions about K2 light curves for some stars, and typical TESS light curves are definitely too short.

\section{Conclusions and Summary}\label{summary}

We have developed a methodology based on simple properties of light curve autocorrelation functions to estimate a lifetime for starspots on stars with effective temperatures between 3500-6800K that have not evolved well past the main sequence. By ``lifetime" we mean the temporal length of the physical presence of coherent spotted regions whose integrated size evolves throughout their presence. To measure this we test the persistence of patterns over several rotations in precision differential broadband light curves having dense and long coverage. Our method provides the first general relationships between starspot lifetimes, stellar rotation periods, and effective temperatures in a large ($>60,000$) sample of stars from the Kepler Prime mission. We study spot lifetimes both in days and in units of the stellar rotation period itself; these produce somewhat different conclusions. The lifetimes probably refer more to starspot groups than individual spots, and could be reflective either of the extended existence of a coherent spot group or the emergence of magnetic flux over an extended period in a fairly large (tens of square degrees) area on a star (we do not have the spatial resolution to distinguish between these). 

The primary diagnostic used in our method is the comparison of correlation strengths over 1-5 rotations in a normalized and properly conditioned autocorrelation function (ACF) from a long light curve segment. This conditioning includes choosing peaks that are for integer harmonics of the rotation period. We find that the large sample of Kepler stars presented by \citet{MMA14} (MMA14) exhibits very systematic behavior in plots of the second or third ACF peak height (strength) against the first peak height. The peak-peak plots (PPPs) from the data are compared to PPPs generated using spot models like those from \citet{Basri20} with a range of spot lifetimes from 1 to 50 stellar rotations. The model PPPs are qualitatively similar to but clearly distinguishable from the observed PPPs. 

We were able to render the model PPPs quite similar to the observations by combining light curves from short lifetime (one rotation) models with models of other longer lifetimes, in ratios that depend on how long the other lifetimes are. In general we needed a smaller mix of short lifetimes as the other lifetimes got longer. This could be interpreted to mean that as stars are increasingly covered by larger longer-lived spot groups, the contribution and influence of short-lived spots becomes weaker. Once we established a set of combined model light curves that generated similar PPPs to the observations, we could back out the contributions of various lifetimes to different regions of the PPP. It turns out that this relation can be expressed very simply as a (logarithmic) polynomial function between the normalized correlation strength for the first rotation (first ACF peak height) and the spot lifetime in rotations.

We tested the effects of stellar inclination and differential rotation on the model PPPs. Very low inclinations and strong differential rotation each have significant effects and tend to shorten the lifetime derived from a particular Pk1. The most reliable result we find is that there are three basic groups of relations between Pk1 and $L_{rot}$ that seem fairly robust against the tests we made. Short rotational lifetimes (1-4) are implied when Pk1 is less than about 0.4, long lifetimes (15 or greater) are implied when Pk1 is greater than about 0.7, and in between are the intermediate cases. The internal uncertainties in $L_{rot}$ are roughly 2-5 for a given value of Pk1 (and Pk3); the larger uncertainties apply to longer lifetimes. Lifetimes between 10-15 rotations show the greatest dependence on stellar inclination and differential rotation in their Pk1 distributions. 

We examined the question of how many rotation periods a light curve needs for one to infer a spot lifetime with reasonable confidence. This depends on spot lifetimes in the sense that it is less important for stars with short spot lifetimes because they will deliver that result regardless. For stars with longer spot lifetimes it is preferable to have something near 50 rotation periods in the light curve, otherwise one may infer shorter than actual lifetimes for some of them. The Kepler Prime dataset is of adequate length. There are definitely also unknown systematic uncertainties because we don't know how well these particular schematic models represent reality. Despite the caveats above we derive a specific relation between Pk1 and $L_{rot}$ from our base case. We then apply the derived relation to the observed Pk1 values, yielding an estimated $L_{rot}$ for each Kepler star. This allows examination of relations between starspot group lifetimes and stellar parameters.

When spot lifetimes become short it is harder to derive a rotation period, so the MMA14 sample is biased towards stars with long enough lifetimes and larger photometric variability. We therefore also analysed a similarly large sample of main sequence Kepler stars for which MMA14 were unable to confidently derive rotation periods (we call this the NP21 sample). Those stars lie almost entirely in the short lifetime region of the PPP. They are probably as old or older than the Sun, which itself lies in this part of the PPP. We derived pseudo-periods for these stars by simply choosing the location of the highest of the first three ACH peaks. This same procedure does an excellent job of reproducing the MMA14 rotation periods, but for the NP21 stars there is a large puzzling excess at pseudo-periods between 10-20 days. These probably represent half-periods, but resolving that dilemma is beyond the scope of this paper. It does not much affect our main conclusions about starspot lifetimes, however, because the NP21 stars definitely have short lifetimes regardless.

There is a general trend for the MMA14 stars to have longer $L_{rot}$ at shorter $P_{rot}$ as expected. There is a also a separate much sparser and diffuse branch of them that have fairly constant $L_{rot}$ below 10 at $P_{rot}$ less than about 10 days. If we assume the NP21 periods should be doubled below about 23 days then the NP21 sample fits well at the long-period short-lifetime end of the MMA14 sample, as well as containing even longer periods with short lifetimes. The NP21 sample is also at the small end of photometric ranges; there is a general tendency for stars with larger ranges to have longer $L_{rot}$ and shorter $P_{rot}$ as has been suggested in previous papers. 

Since the MMA14 stars all have derived rotation periods and we have pseudo-periods for the NP21 sample, we also convert spot lifetimes in rotation periods to physical lifetimes in days. The overall distribution of $L_{day}$ is lognormal for both stellar samples. The distribution for the MMA14 sample peaks at a lifetime around 140 days, while the NP21 sample peaks around 80 days. It is the shorter spot lifetimes of the NP21 stars that makes their rotation periods harder to determine. The appearance of the $L_{day}$-$P_{rot}$ plot is different from the $L_{rot}$-$P_{rot}$ plot because stars can move around quite a bit after multiplying $L_{rot}$ by $P_{rot}$. The shorter period stars, for example, actually have shorter $L_{day}$ and longer $L_{rot}$ at a given photometric range than longer period stars because the influence of $P_{rot}$ is stronger for them.

Most Kepler stars now have Gaia stellar parameters so we can examine the relation of spot lifetime to stellar effective temperature. The general effects seen are that $L_{rot}$ is not very dependent on $T_{\rm eff}$ but $L_{day}$ increases as temperature decreases. This latter effect is influenced by the fact that $P_{rot}$ exhibits the same behavior. We divided the stars into 8 temperature groups in bands of 400K. The 6600K group shows generally fast rotation and short lifetimes, although longer lifetimes at short periods already appear at 6200K. There is a surprising but small group of hot stars with quite short lifetimes at quite short rotation periods. Some of those periods may be due to pulsations rather than starspots, or spots may be relatively short-lived in very thin convection zones. 
The solar-type stars show the general relations described above most clearly since most of the Kepler stars fall into this category and so influence the overall behavior most. The general appearance of the $L_{day}$-$T_{\rm eff}$ diagram is quite similar to the $P_{rot}$-$T_{\rm eff}$ diagram, reflecting the relations between these three parameters. The cool stars have a larger population at longer rotation periods, but also show larger scatter in $L_{day}$ leading to more long values of $L_{day}$. The fraction of stars above the bulk of the values for $L_{day}$ (about 200 days) is a clearly increasing function of decreasing $T_{\rm eff}$. The ratio of stars above to stars below this line in $L_{day}$ for all rotation periods goes from 0.025 for the hottest group to 1.11 for the coolest group. This seems to imply that spot group lifetimes grow longer as the convection zone deepens.

In summary, this paper has presented a new methodology for extracting information from light curves using autocorrelation functions to study the persistence of patterns in light curves due to starspot evolution. Rather than considering the ``decay" of the ACF it examines the relations between the strengths of ACF peaks as a function of starspot lifetime. It can no doubt be improved upon, and more detailed analyses could be conducted. In particular it will be interesting to try calibrating the method with models that generate light curves using different and more physical assumptions and parameters, different constraints on starspot distributions and differential rotation, or even different interpretations of ``lifetime". What seems clear is that examining the relations between the first few autocorrelation peak strengths is a productive approach to extracting information on starspot evolution timescales from precision light curves with enough temporal coverage.

\begin{acknowledgments}
The Kepler mission was absolutely critical to this work. We thank NASA for approving the mission, the Kepler team for conducting a successful mission, and the MAST for making its data generally available. GB enjoyed fruitful discussions with Drs. Sami Solanki, Alexander Shapiro, Timo Reinhold, Nina N{\`e}mec, and Emre I{\c s}{\i k} at MPI G{\"o}ttingen.
\end{acknowledgments}

{\it Facilities:} \facility{Kepler}, \facility{MAST}.

\appendix

\end{document}